\documentclass[apj]{emulateapj}
\usepackage{ulem,color,epstopdf} 
\usepackage{amsmath, graphicx}

\usepackage{hyphenat}
\def\kw{{Konus-\textit{Wind}}}
\def\KW{{KONUS-\textit{WIND}}}
\def\mw{{microwave}}
\def\Mw{{Microwave}}

\def\p{_{\rm peak}}
\def\csf{{cold solar flare}}
\def\eif{{early impulsive flare}}

\def\gyr{{gyrosynchotron}}

\received{2017/10/26}

\revised{2017/11/17}

\accepted{2018/02/22}

\begin{document}

\title{Statistics of ``Cold''  Early Impulsive Solar Flares in X-ray and Microwave domains}

\author{Alexandra L. Lysenko}
\affiliation{Ioffe Institute, Polytekhnicheskaya, 26, St. Petersburg, 194021 -  Russia}

\email{alexandra.lysenko@mail.ioffe.ru}

\author{Alexander T. Altyntsev}
\affiliation{Institute of Solar-Terrestrial Physics  (ISZF), Lermontov st., 126a, Irkutsk, 664033 -  Russia}

\author{Natalia S. Meshalkina}
\affiliation{Institute of Solar-Terrestrial Physics  (ISZF), Lermontov st., 126a, Irkutsk, 664033 -  Russia}

\author{Dmitriy Zhdanov}
\affiliation{Institute of Solar-Terrestrial Physics  (ISZF), Lermontov st., 126a, Irkutsk, 664033 -  Russia}

\author{Gregory D. Fleishman}
\affiliation{New Jersey Institute of Technology, University Heights, Newark, NJ 07102-1982 -  United States}
\affiliation{Ioffe Institute, Polytekhnicheskaya, 26, St. Petersburg, 194021 -  Russia}

\begin{abstract}

Solar flares often happen after a preflare / preheating phase, which is almost or entirely thermal. In contrast, there are the so-called \eif s that do not show a (significant) preflare heating but instead often show the Neupert effect--a relationship where the impulsive phase is followed by a gradual, cumulative-like, thermal response. This has been interpreted as a dominance of nonthermal energy release at the impulsive phase, even though a similar phenomenology is expected if the thermal and nonthermal energies are released in comparable amounts at the impulsive phase. Nevertheless, some flares do show a good quantitative correspondence between the nonthermal electron energy input and plasma heating; in such cases the thermal response was weak, which results in calling them ``cold'' flares. We undertook a systematic search of such events among  \eif s registered by \kw\ instrument in the triggered mode from 11/1994 to 04/2017 and selected 27 cold flares based on relationships between HXR (\kw) and SXR (GOES) emission. For these events we put together all available microwave data from different instruments. We obtained temporal and spectral parameters of HXR and microwave emissions of the events and examined correlations between them. We found that, compared with  a `mean' flare, the cold flares: (i) are weaker, shorter, and harder in the X-ray domain; (ii) are harder and shorter, but not weaker in the \mw s; (iii) have a significantly higher spectral peak frequencies in the \mw s. We discuss the possible physical reasons for these distinctions and implication of the finding.

\end{abstract}


\section{Introduction}

Explosive energy release that results in efficient acceleration of charged particles and heating of the ambient plasma is ubiquitous in the astrophysics. In the solar atmosphere this energy release is observed as transient brightening at various wavelengths, most notably during solar flares. In the solar flares the excessive magnetic energy can be promptly released at a time scale as short as seconds, while generating nonthermal particles with high energies as large as 1~MeV  and heating the ambient plasma up to several dozens million Kelvin. Although particle acceleration and plasma heating are typical for many astrophysical objects, the solar flares offer a unique natural laboratory, where these processes can be studied at extremely short spatial and temporal scales needed to address the key dynamics of the explosive energy release. It is puzzling that the proportions of the energy that initially go to either particle acceleration or plasma heating vary dramatically between different solar flares. Indeed, there are entirely thermal flares \citep{Gary_Hurford_1989, Battaglia_etal_2009, Fl_etal_2015}, where no nonthermal emission is detected, while in some flares an exceptionally strong nonthermal emission is accompanied by a very modest thermal component \citep{White1992,Bastian_etal_2007, Fl_etal_2011, Masuda2013}. Not surprisingly, there are cases with all possible proportions between these mentioned extremes of purely thermal and entirely nonthermal flares.

It is not yet clear what physical process or parameter combination is decisive for the initial energy partitions in the flares, neither what is the fundamental difference between a `normal' solar flare, in which the thermal and nonthermal energies are initially comparable, and those dominated by the nonthermal component. In this study we are going to advance these questions using a statistical comparison between events from various subgroups described below.
There are numerous statistical studies of solar flares in the X-ray and \mw\ domains that employ data from many past or existing instruments. In particular, \cite{Tanaka1983}, based on HINOTORI data, and \cite{Dennis1985}, based on SMM/HXRBS data, performed the statistical analysis of the X-Ray bursts registered during solar cycle 21 and revealed three types of X-Ray solar flares: the first type represents soft hot flares with minor HXR emission, the flares of the second type have noticeable HXR emission, impulsive time profiles and soft-hard-soft spectral evolution with typical power-law spectral indices between $\sim$2 and $\sim$8, and the flares of the third type are characterized by extended durations, soft-hard-harder spectral evolution and power-law indices $\leq$4.5 \citep{BaiSturrock1989}. \cite{Dennis1985} also established that the occurrence rate distribution of the HXR peak fluxes obeys a power-law with index -1.8. Later, power-laws were also found for distributions of HXR fluences with power-law indices between -1.53 and -1.77 and HXR durations with power-law indices between -1.76 and -2.54 \citep{Crosby1993, Aschwanden2011}.
The statistical comparison between HXR and microwave flare components performed by \cite{Kosugi1988} shows, that HXR emission in impulsive flares at $\sim$100~keV and \mw\ emission at 17~GHz are highly correlated and presumably are the results of nonrelativistic down-streaming electrons, while in extended flares \mw s are emitted by relativistic electrons trapped in coronal loops. \cite{Silva2000} performed the analysis of correlations between paramaters of HXR burts registered by CGRO/BATSE and \mw\ by OVSA, this analysis also revealed that HXR fluxes in $\leq$200~keV and \mw\ fluxes are generally correlated, while HXR and \mw\ spectral parameters are often unrelated. One of the most detailed statistical studies in the \mw\ domain was performed by \cite{Nita_etal_2004} for flares registered by the Owens Valley Solar Array (OVSA) during 2001--2002.

\cite{Su2008}, based on the RHESSI data, introduced three solar flare types according to their relationship between thermal and nonthermal emission. Type 1, ``accordantly gradual flares'', are purely thermal with no obvious emission above 25~keV. Type 2 flares, called ``accordantly impulsive flares'', demonstrate impulsive nonthermal emission in HXRs along with more gradual thermal SXR emission. In the case of type 3 flares, ``early impulsive flares'', first proposed by \cite{Sui2007}, impulsive nonthermal phase is followed by thermal emission. In a subset of impulsive flares, a so-called Neupert effect is observed \citep{Neupert1968}. Originally, \citet{Neupert1968} discovered a correlation between the flare time profiles in SXR, where the thermal emission is observed, and the cumulative integral of the time profile in \mw s, where the emission is generated by nonthermal particles. Later, a similar relationship was revealed between SXRs and HXRs. Taken at the face value, the Neupert effect implies that the thermal plasma is heated by nonthermal electrons. However, this heating by the nonthermal electrons might not be the only heating occurring during the flare. Indeed, if the thermal plasma is somehow impulsively heated directly by the magnetic energy release at the same time scale as that of the electron acceleration, the light curves of thermal emission will still demonstrate the Neupert effect because the thermal plasma cooling is a much slower process than the impulsive energy release. Such additional heating was proposed by \citet{Veronig2005} in a subset of events that demonstrate the phenomenology of the Neupert effect. In additions, other ways of plasma heating are  observed or implied in solar flares, so the Neupert effect is not always observed \citep{Su2008, Battaglia_etal_2009}.

Although in a general case the presence of the Neupert effect does not guarantee that the plasma heating is solely driven by the nonthermal electrons, it would not be unreasonable to expect that there are flares in which this heating mechanism does dominate. Most likely, such flare would represent a subset of \eif s, whose thermal emission is very low before the impulsive phase, so the thermal plasma is likely heated by almost only nonthermal particles.
Such cases, called ``cold'' solar flares (CSFs hereafter), that show very low thermal response relative to nonthermal energy of accelerated particles, have been reported in previous case studies \citep{White1992, Bastian_etal_2007, Fl_etal_2011, Masuda2013}. However, the criteria to classify a given flare as a ``cold'' one were somewhat subjective; typically, the reported CSFs showed a noticeable HXR and/or \mw\ burst that happened without any reported GOES flare. Yet, no formal criterion has  been developed of how weak the thermal response must be compared to the nonthermal emission for the flare to be classified as a cold one. Accordingly, with all the variety of statistical studies described above, no focused statistical study on the flares with weak thermal response, the ``cold'' flares, has yet been available.

It is important to realize that there are two formal reasons to render the thermal SXR emission low: (a) low plasma temperature \citep[as in the cases reported by][]{Bastian_etal_2007, Masuda2013} or (b) low emission measure due to either a low plasma density \citep[as in the tenuous flare reported by][]{Fl_etal_2011} or a small volume of the flaring loop \citep[as in the main flaring loop in the flare reported by][]{Fl_etal_2016}\footnote{In this flare the thermal response was very weak at the impulsive phase because a small loop with a correspondingly small volume produced the impulsive HXR and \mw\ emission. However, one more, much bigger loop was also involved in the flaring, which was responsible for a delayed thermal response in this, rather unusual event.}; thus, in case (b) the observed thermal response can remain low even if the plasma temperature is high. With this reservation, here we will use the term ``cold flare'' for any event with a reasonably weak thermal signature (see below for formal criteria), because it is not at all easy to sort out cases (a) and (b) at the stage of the event selection.

Here we take advantage of availability of an almost uniform database of solar flares recorded by the \kw\ \citep{Aptekar1995} during two solar cycles (between November, 1994 and April, 2017) to build a statistically significant subset of the cold flares and study their properties as compared to other flares. 
Based on the performed statistical analysis, we discuss what flare properties or parameter combination make the CSF different from the `normal' flare and what are the likely main causes of the apparent lack of thermal emission in this class of solar flares.


\section{Instrumentation and Event Selection}

\subsection{\label{X-ray} X-ray Domain: Uniform Input from the \KW\ and the GOES}

Given that CSFs occur  relatively seldom, the statistical study of these events requires a reasonably long series of observations. From this perspective we employ hard X-ray (HXR) data from the \kw\ and soft X-ray (SXR) data from the GOES; both data sets are available over the time period longer than two full solar cycles \citep{White2005}.

\subsubsection{\label{S_GOES} GOES Soft X-ray Data}

Spacecrafts of GOES (Geostationary Operational Environmental Satellite) series observe the Sun almost continuously since 1974. We used data of GOES X-ray sensors in two broadband SXR channels, softer channel 1--8~\AA\ and harder channel 0.5--4~\AA , with temporal resolution varied from 3~s to 2.046~s during observational history.

\subsubsection{\label{S_KW} \KW\ Hard X-ray Solar Data}

\kw\ was launched on 1 November 1994 on board of the Wind spacecraft to detect gamma-ray bursts and solar flares in HXR domain. It operates in interplanetary space and since July  2004 is located near Lagrange point L1 at $\sim$5~light seconds from the Earth. \kw\ consists of two identical 13~cm~x~7.5~cm NaI(Tl) detectors S1 and S2 with Be entrance window. Detectors are located on the opposite sides of the spacecraft observing the southern and the northern ecliptical hemispheres correspondingly \citep{Aptekar1995}.

The \kw\ works in two modes---the waiting mode and the triggered mode. In the waiting mode, only the light curves with accumulation time 2.944 s are available in three wide energy bands: G1 (nominal range 13--50~keV), G2 (nominal range 50--200~keV), G3 (nominal range 200--750~keV). The \kw\ energy boundaries changed during its observational history within a factor of 2.0 for detector S1 and 1.5 for detector S2.

Switching to the triggered mode\footnote{Light curves and spectra of solar flares registered by \kw\ in the triggered mode can be found at \url{http://www.ioffe.ru/LEA/kwsun/}} occurs at a statistically significant background excess on 1~s or 140~ms timescale in energy band G2. In the triggered mode, the light curves are recorded in the same three energy bands with the high time resolution (varying from 2 to 256 ms as the burst progresses) during 230 s along with accumulation of 64 multichannel energy spectra. Multichannel spectra are measured in partially overlapping energy ranges: nominal boundaries are 13--750~keV for the first range and 250~keV--15~MeV for the second range. Now the spectral ranges have changed to $\sim$25~keV--18~MeV for the S1 detector and $\sim$20~keV--15~MeV for the S2 detector from the original 13~keV--10~MeV. Each energy range consists of 63 energy channels. Accumulation time for each of the first 4 spectra is 64~ms, while the accumulation time varies for the subsequent 52 spectra from 256~ms to 8.192~s according to the count rate in G2 energy band: for stronger HXR flux the accumulation time is proportionally shorter. For the latest eight spectra the accumulation time is fixed at 8.192~s.  When accumulation of the triggered mode light curves and energy spectra has been completed, both triggered and waiting mode measurements are interrupted by a gap of $\sim$1~hour because of the data readout.

Due to adopted implementation of the trigger algorithm, in the triggered mode the \kw\ registers only reasonably hard flares showing a rather rapid increase in HXR intensity,  while for softer and smoother events only waiting mode data are available. Because of the limited duration of the  trigger record ($\sim$240~s for time history and $\sim$480~s maximum for energy spectra) for longer flares the recording ends before the end of the flare \citep{Pal'shin2014}.

Given the lack of spatial resolution, an  algorithm is needed to distinguish solar bursts from other astrophysical sources. There is a number of criteria to conclude if a \kw\ triggered mode event is a solar flare or not. First, as \kw\ detectors are pointed transversely to ecliptic plane, emission from solar flares reaches \kw\ at angles $\sim$90$^\circ$ and is seen in both detectors S1 and S2---in one detector in the triggered mode, while in the other---in the waiting mode (the trigger may only occur in one detector at a time). Second, we check the GOES X-ray event list\footnote{GOES event list \url{ftp://ftp.swpc.noaa.gov/pub/indices/events/}} for event notification or look for increase in the GOES SXR flux at that time. Also the Fermi trigger reports may be used\footnote{Fermi trigger information \url{https://gcn.gsfc.nasa.gov/fermi\_grbs.html/}} for a subset of jointly observed events.

\begin{deluxetable*}{llllcc}[t!]
\tablecolumns{6}
\tablewidth{0pc}
\tablecaption{\label{table_KW_flare_list} \kw\ \csf\ list.
}
\tablehead{\colhead{N} & \colhead{strID\tablenotemark{*}}  & \colhead{Date}	& \colhead{$t_0$\tablenotemark{**}, hh:mm:ss} & \colhead{Coordinates} & \colhead{Instrument\tablenotemark{***}}
}
\startdata
1 &	SOL1998-05-07T19952 &	1998-May-07 & 05:32:37.072 & [554$\arcsec$, 495$\arcsec$] & EIT, 195 \AA\ \\
2 &	SOL1999-06-19T82485 &	1999-Jun-19 & 22:54:49.788 & [-916$\arcsec$, 233$\arcsec$] & NoRH \\
3 &	SOL1999-07-30T82686 &	1999-Jul-30 & 22:58:09.675 & [-490$\arcsec$, 320$\arcsec$] & NoRH \\
4 &	SOL1999-11-09T30381 &	1999-Nov-09 & 08:26:21.703 & \nodata & \nodata \\
5 &	SOL1999-11-14T53708 &	1999-Nov-14 & 14:55:08.244 & \nodata & \nodata \\
6 &	SOL1999-12-02T72060 &	1999-Dec-02 & 20:01:00.012 & \nodata & \nodata \\
7 &	SOL2000-03-10T15704 &	2000-Mar-10 & 04:21:48.688 & [-776$\arcsec$, -211$\arcsec$] & NoRH \\
8 &	SOL2000-03-18T08707 &	2000-Mar-18 & 02:25:10.567 & [717$\arcsec$, -294$\arcsec$] & SSRT \\
9 &	SOL2000-05-18T26517 &	2000-May-18 & 07:21:59.706 & [-33$\arcsec$, -257$\arcsec$] & SSRT \\
10 &	SOL2000-05-18T82777 &	2000-May-18 & 22:59:39.777 & [177$\arcsec$, -319$\arcsec$] & NoRH \\
11 &	SOL2001-10-12T27630 &	2001-Oct-12 & 07:40:31.941 & [-900$\arcsec$, 238$\arcsec$] & SSRT \\
12 &	SOL2001-11-01T55062 &	2001-Nov-01 & 15:17:42.772 & [320$\arcsec$, 125$\arcsec$] & OVSA\\
13 &	SOL2002-05-29T27586 &	2002-May-29 & 07:39:46.864 & [-343$\arcsec$, 122$\arcsec$] & SSRT \\
14 &	SOL2002-08-10T85808 &	2002-Aug-10 & 23:50:09.293 & [-938$\arcsec$, -79$\arcsec$] & NoRH \\
15 &	SOL2002-08-18T83478 &	2002-Aug-18 & 23:11:19.740 & [300$\arcsec$, -300$\arcsec$] & OVSA \\
16 &	SOL2002-08-20T71727 &	2002-Aug-20 & 19:55:28.476 & \nodata & \nodata \\
17 &	SOL2003-10-23T80262 &	2003-Oct-23 & 22:17:39.620 & [-938$\arcsec$, -307$\arcsec$] & RHESSI \\
18 &	SOL2005-09-08T08145 &	2005-Sep-08 & 02:15:49.996 & \nodata & \nodata \\
19 &	SOL2011-09-19T27816 &	2011-Sep-19 & 07:43:40.791 & [-806$\arcsec$, 345$\arcsec$] & RHESSI \\
20 &	SOL2012-07-08T09826 &	2012-Jul-08 & 02:43:50.647 & [894$\arcsec$, -206$\arcsec$] & NoRH \\
21 &	SOL2013-11-05T13819 &	2013-Nov-05 & 03:50:24.588 & [-771$\arcsec$, -250$\arcsec$] & NoRH \\
22 &	SOL2014-01-02T20697 &	2014-Jan-02 & 05:45:01.390 & [-948$\arcsec$, -83$\arcsec$] & NoRH \\
23 &	SOL2014-01-31T60753 &	2014-Jan-31 & 16:52:37.461 & [-504$\arcsec$, 331$\arcsec$] & RHESSI \\
24 &	SOL2014-02-08T20965 &	2014-Feb-08 & 05:49:29.848 & [856$\arcsec$, -150$\arcsec$] & RHESSI \\
25 &    SOL2014-10-18T10152 &	2014-Oct-18 & 02:49:17.710 & [-909$\arcsec$, -335$\arcsec$] & RHESSI \\
26 &    SOL2014-10-27T11681 &   2014-Oct-27 & 03:14:46.862 & [640$\arcsec$, -290$\arcsec$] & RHESSI \\
27 &    SOL2015-05-07T45695 &   2015-May-07 & 12:41:40.415 & \nodata & \nodata \\
\enddata
\tablenotetext{*}{Here T***** is the \kw\ trigger time in seconds without time of light propagation corrections to match the format used in \url{http://www.ioffe.ru/LEA/kwsun/}.}
\tablenotetext{**}{The \kw\ trigger time after corrections for the light propagation to the Earth are applied.}
\tablenotetext{***}{The instrument used for the event localization.}
\mbox{}
\end{deluxetable*}

\begin{figure*}\centering
\includegraphics[width=0.95\textwidth]{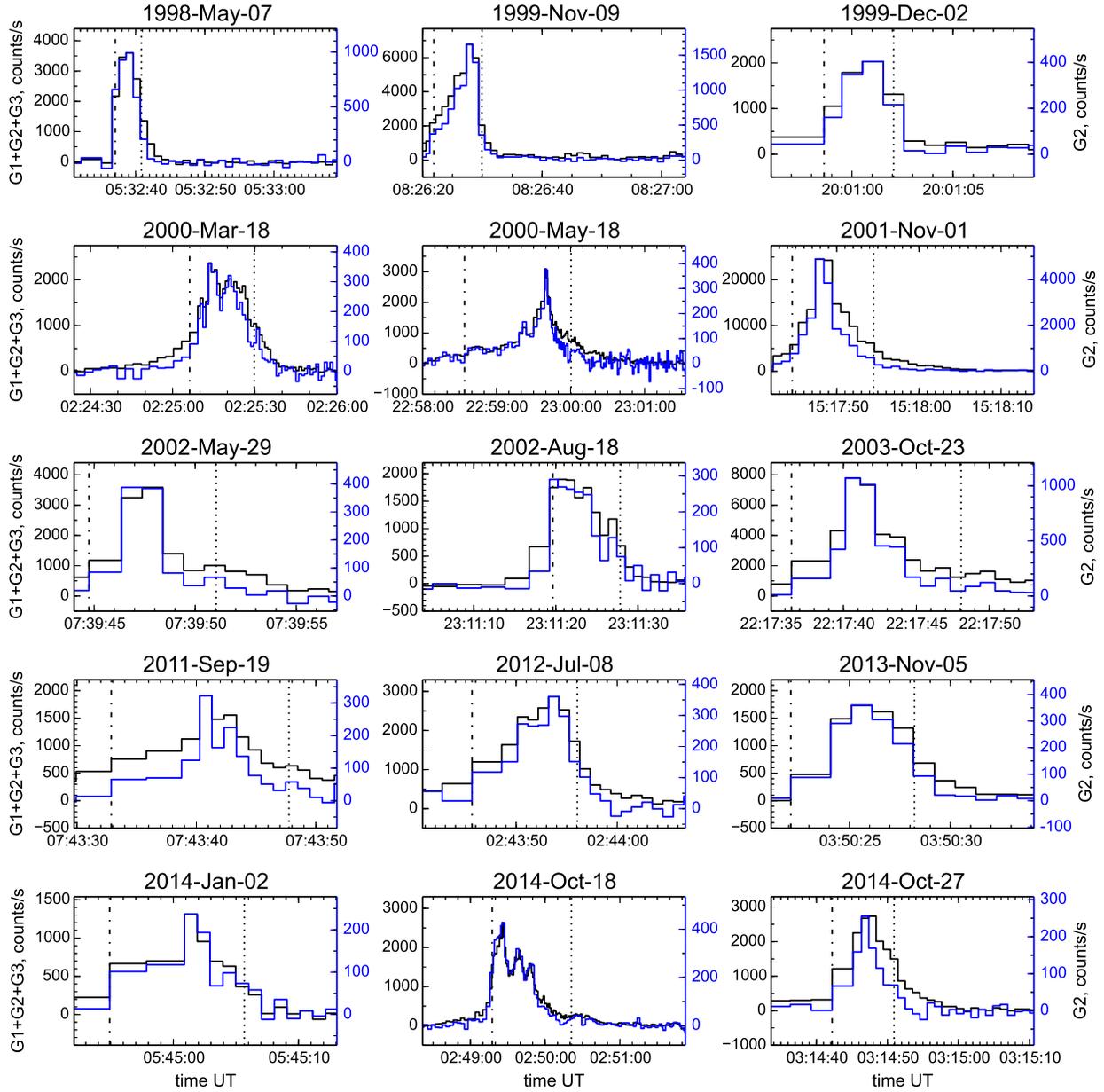}
\caption{\label{kw_curve_mw_fit} CSF time profiles in HXR range measured by \kw\ in the sum G1+G2+G3 channels ($\sim$20--1200~keV, black curve, left panel) and in G2 channel ($\sim$80--300~keV, blue curve, right panel). For the ease of comparison with Figure~\ref{MW_dyn_fit}, here we show 15 cases for which the \mw\ data allowed spectral fitting.}
\end{figure*}

\subsubsection{\label{EIF_selection} Selection of Early Impulsive Flares }

To form an initial list of the event candidates, we employed the following formal criterion for the ``solar flare''-like burst registered by \kw\ in the triggered mode to be listed as \eif : we require that no GOES X-ray event has been reported  at the time of the \kw\ trigger. This means that either the corresponding GOES event began later than the \kw\ trigger time or there was no reported GOES event at all. The goal of using that strict criterion is to  exclude events with SXR emission due to plasma preheating by any other agent\sout{s} than the nonthermal flare-accelerated electrons.

This automatic search yielded 84 events. Some of these events were then discarded manually because of failures in the \kw\ data, some events were identified as false alarms caused by energetic particles, not HXR emission. Three events are missing from the GOES event list for an unknown reason, though they demonstrate noticeable increase of GOES 1--8 \AA\ flux before the HXR impulsive phase, and, thus, do not obey the entry criterion. Two event were discarded due to failures in the GOES data.

\begin{figure*}\centering
\includegraphics[width=0.95\textwidth]{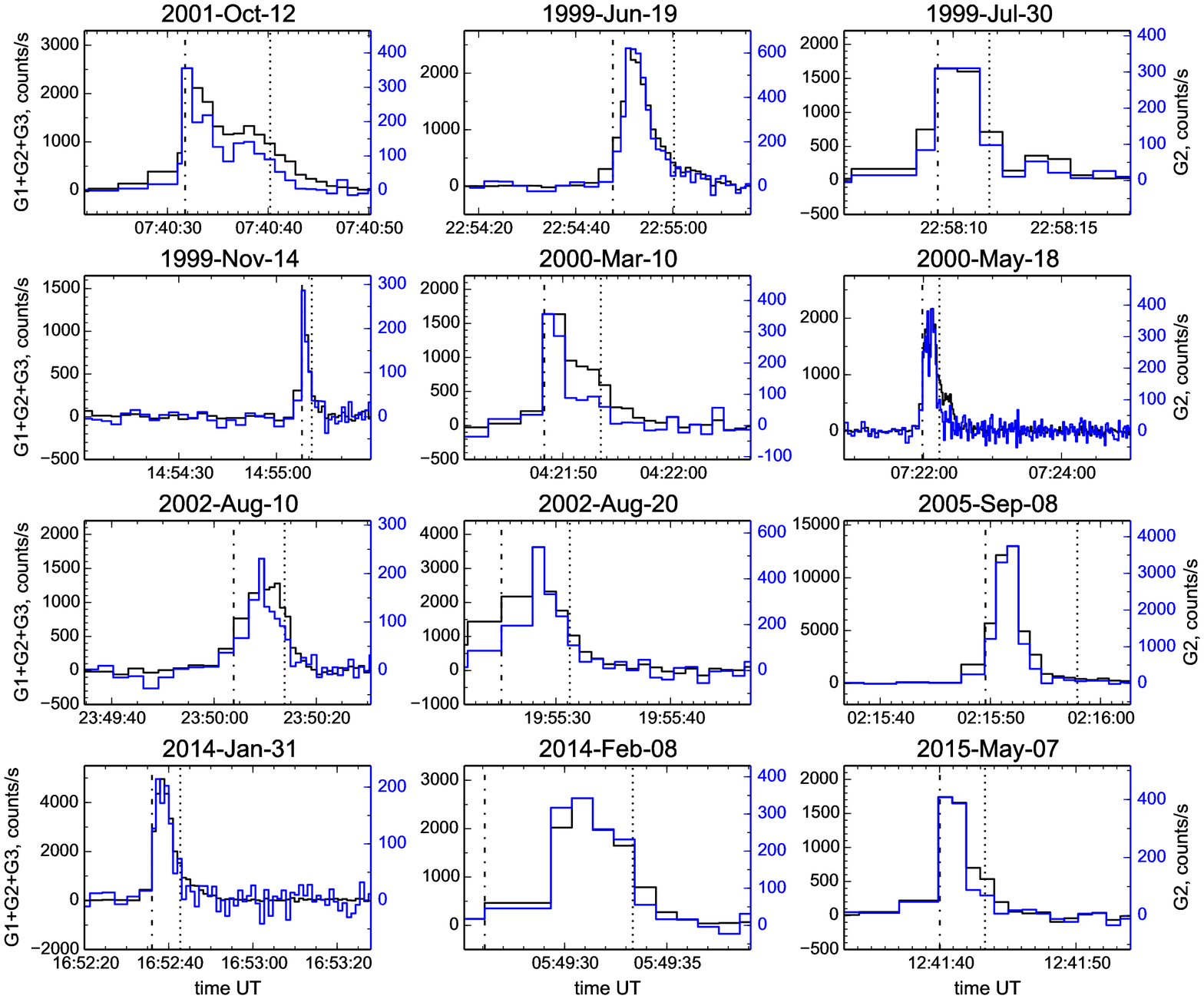}
\caption{\label{kw_curve_mw_nofit} CSF time profiles in HXR range measured by \kw\ in the sum G1+G2+G3 channels ($\sim$20--1200~keV, black curve, left panel) and in G2 channel ($\sim$80--300~keV, blue curve, right panel). For the ease of comparison, here we show 12 cases corresponding to Figure~\ref{MW_dyn_nofit}.}
\end{figure*}

Finally, we approved 42 solar flares, whose properties are consistent with those of \eif s proposed by \cite{Sui2007}: no increase in SXR flux must be seen earlier than 30~s prior to the increase in the HXR flux. For all forty two events there is no corresponding reported GOES flare. Thus, our criterion for the \eif\ selection is really rather strict. The absence of a solar flare in the GOES event implies that it did not obey the adopted GOES flare selection criterion: a solar flare is listed as a GOES flare if it demonstrates monotonous flux increase in the GOES 1--8~\AA\ channel during at least 1~minute as compared to previous 3~minutes. Thus, short events and events during unstable background could be missed from this list. In what follows, see Section~\ref{HXR_vs_SXR}, based on cross-correlation analysis between HXR and SXR data, we selected only 27 cold out of these 42 \eif s for a more detailed analysis, listed in Table~\ref{table_KW_flare_list}. \kw\ time profiles for these events are presented in Figures~\ref{kw_curve_mw_fit} and \ref{kw_curve_mw_nofit}.


\subsubsection{\label{S_ref_selection} Selection of Reference Flares }

Because of the trigger nature of the \kw\ instrument, which records only reasonably hard impulsive HXR bursts (see Section~\ref{S_KW} above), we could not make a meaningful direct comparison of our cold flare subset with previously available statistical studies. This forced us to form and use a reference set of other bursts recorded by the \kw\ in the triggered mode, with which the CSFs are to be compared. Accordingly, from all 1000+ solar flares recorded by \kw\ in the triggered mode, for the reference set we selected flares that (i) have constant \kw\ and GOES background and (ii) fully covered by the \kw\ time history record (both in the waiting and in the triggered mode), i. e. which ended before the end of the trigger record (but may begin in the waiting mode record before the trigger mode record begin). This last condition implies discarding long-duration flares. In this way we selected 405 C, M, and X GOES class flares to form the reference set.

\subsection{\label{MW_range}\Mw\ domain: nonuniform input}

Historically, the microwave data played a primary role in identification and analysis of the cold flares \citep{White1992, Bastian_etal_2007, Fl_etal_2011, Masuda2013, Fl_etal_2016}; thus, we have to  use all available microwave data fully. Unfortunately the only quasi-uniform set of radio instruments is Radio Solar Telescope Network \citep[RSTN,][]{Guidice1981}, which has a lot of disadvantages including clock errors, calibration errors, big gaps between the working frequencies, and a limited spectral coverage. For this reason, in addition to RSTN, we use several other radio instruments; namely the Owens Valley Solar Array \citep[OVSA,][]{ovsa_1984, Gary_Hurford_1994}, the Nobeyama Radio Polarimeters \citep[NoRP,][]{Torii_etal_1979}, the Solar Radio Spectropolarimeters \citep[SRS,][]{Muratov2011}, the Badary Broadband Microwave Spectropolarimeters \citep[BBMS,][]{Zhdanov2015}, and the Kislovodsk Mountain Astronomical Station of the Pulkovo Observatory \citep[KMAS,][]{Shramko2011}; see Table~\ref{table_MW_instruments}. This approach, however, has a disadvantage of making the data input nonuniform. We will discuss implications of this nonuniformity later.

\begin{deluxetable}{llccl}
\tabletypesize{\small}
\tablecolumns{5}
\tablewidth{0pc}
\tablecaption{\label{table_MW_instruments}Specifications of the instruments used for the \Mw\ database.
}
\tablehead{\colhead{Instr.} & \colhead{Frequencies, GHz} & \colhead{Obs. Time, UT} & \colhead{Time Res.,} \\
\colhead{} & \colhead{} & \colhead{} & \colhead{High/Low, s}
}
\startdata
OVSA &	1.2--18; 40 Channels  &	$\sim$16:00--24:00 &	$4/8\tablenotemark{*}$ \\
NoRP &	1, 2, 3.75,  &	$\sim$23:00--07:00 &	0.1/1 \\
 &	9.4, 17, 35, 80\tablenotemark{**}  &	 &	 \\
SRS &	 2--24; 16 Channels  &	$\sim$00:00--10:00 &	1.6/1.6 \\
BBMS &	 4--8; 26 Channels  &	$\sim$00:00--10:00 &	0.01/0.01 \\
RSTN &	0.6, 1.4, 2.7, &	24h &	1/1 \\
  &	4.995, 8.8, 15.4  &	 &	 \\
KMAS & 6.1, 9.0 & $\sim$08:00--20:00 & 1/1 \\
\enddata
\tablenotetext{*}{During dedicated campaigns the the OVSA time resolution was higher at the expense of reducing the number of frequency channels.  }
\tablenotetext{**}{Data at 80~GHz are unavailable in the background mode.}
\end{deluxetable}

\subsubsection{\label{MW_instr} Radio Instruments}

RSTN provides radio data with 1 second temporal resolution taken at 8 selected frequencies (245 MHz, 410 MHz, 610 MHz, 1415 MHz, 2695 MHz, 4995 MHz, 8800 MHz, 15400 MHz) from Learmonth, Australia and San Vito, Italy stations.

The OVSA is a solar-dedicated \mw\ array that consisted of 2 27-m and 3--5 2-m antennas at various epoches. The OVSA observed at about 40 frequencies distributed roughly logarithmically between 1.2 and 18~GHz. The OVSA observed those 40 frequencies sequentially over 4~s by small antennas (8~s by big antennas), i.e., 0.1~s per frequency, and can trade-off between the time and spectral resolution depending on the selected observing mode. In a standard mode all frequencies are being observed so the standard time resolution is 4/8~s. The OVSA provided total power data in intensity and circular polarizations and offered a limited ability of the source imaging.

NoRP observe the intensity and circular polarization at six frequencies (1, 2, 3.75, 9.4, 17, \& 35~GHz) and the intensity only at 80~GHz with the time resolution 0.1~sec in the flare mode,  and 1~sec in the background mode (no 80~GHz data).

SRS and BBMS are spectropolarimeters supporting science with Siberian Solar Radio Telescope. SRS measures the integrated flux over the whole solar disk in 2--24~GHz frequency range in two circular polarizations at 16 frequencies with a temporal resolution of 1.6 sec \citep{Muratov2011}. BBMS is the 4--8 GHz spectropolarimeter which measures the integrated flux over the whole solar disk in two circular polarizations at 26 frequencies with a temporal resolution of 10 ms. \citep{Zhdanov2011}.

KMAS measure the integrated solar flux at two frequencies, 6.1~GHz and 9.0~GHz, with time resolution of 1~s. No polarization measurements is available from KMAS.

\subsubsection{\label{MW_database} Building the Microwave Burst Database}

The most comprehensive study of the solar \mw\ burst spectral properties has been performed using the OVSA database accumulated over complete two years of observations during 2001--2002 \citep{Nita_etal_2004}, so it would be beneficial to use a similar database here to make fair comparison of our subset of the data with the statistical distributions of all bursts reported by \citet{Nita_etal_2004}. Unfortunately, only very few events from our list have \mw\ OVSA data. Nevertheless, we made all possible steps to prepare all available data from other radio instruments in a form as similar as possible to the OVSA data; see Appendix~\ref{S_mw_database} for the details. In particular, we combined data obtained by various instruments, fixed clock errors and amplitude calibration errors as well as addressed dissimilar time resolution of the various instruments.

Finally, we built a \mw\ database composed of the 26 events out of the 27 events in the \kw\ list; for  the remaining event no \mw\ data were available. \Mw\ instruments available for each event are listed in Table~\ref{table_mw}.

\subsection{ Cold Early Impulsive Flare Localization}
\begin{figure}\centering
\includegraphics[width=0.5\textwidth]{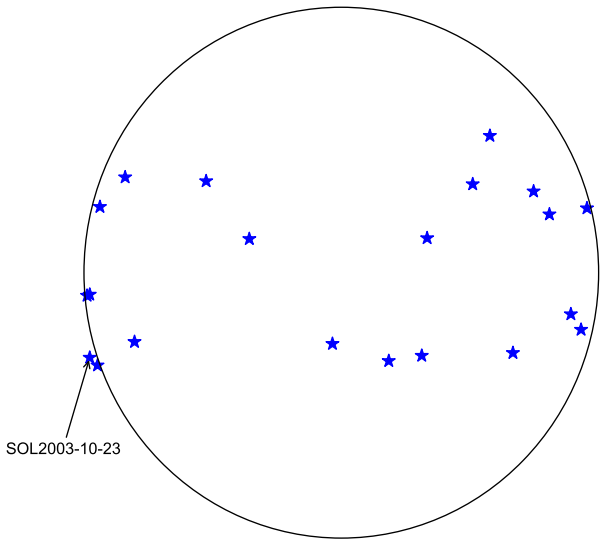}
\caption{\label{fig_loc} CSF positions.}
\end{figure}

We used available imaging information from various instruments to determine flare positions.  Whenever available, we used instruments in X-ray and \mw\ ranges including OVSA, 
Siberian Solar Radio Telescope \citep{SSRT}, Nobeyama RadioHeliograph \citep[NoRH][]{Nakajima1994}, and Reuven Ramaty High Energy Solar Spectroscopic Imager \citep[RHESSI][]{Lin2002}. For one flare, 1998-May-07, we used brightening in the SoHO/EIT \citep{soho}  195~\AA\ data, although this identification may not be reliable, because it corresponds to thermal emission, which is low for CSFs. Eventually, we determined locations of 21 of 27 CSFs, while for the remaining 6 flares no relevant spatial information was found. The heliocentric coordinates of the flares and instruments used for their localization are listed in Table~\ref{table_KW_flare_list}. Locations of the flare are also illustrated in Figure~\ref{fig_loc}, which shows that most flares are located on the solar disk, four flares are near the very limb, and one flare, 2003-Oct-23, demonstrates a source above solar limb.

\begin{figure}[b!]\centering
\includegraphics[width=0.5\textwidth]{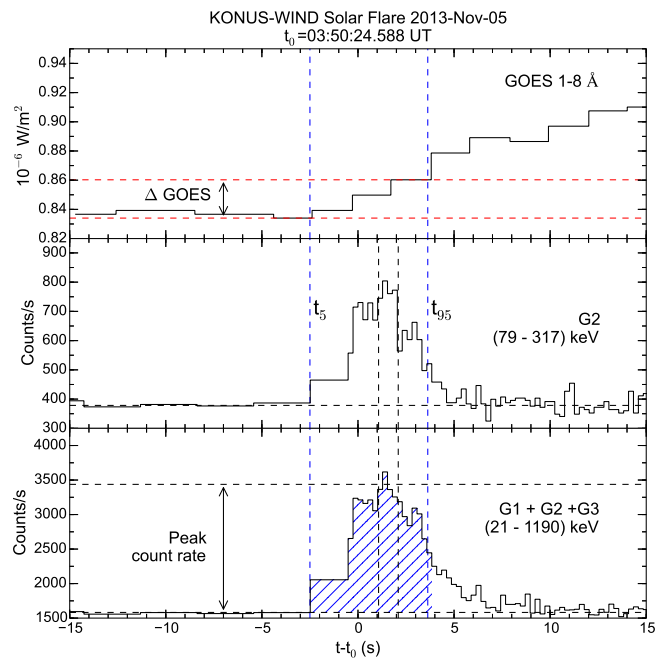}\\
\caption{\label{DeltaGOES_ex} Illustration  for the measurement of the GOES flux increase, $\Delta$GOES, during the impulsive HXR emission. $t_0$ is the \kw\ trigger time after correction for the propagation time from the Wind to the Earth. Vertical dashed blue lines display $t_{90}$, burst duration estimation, determined using  G2 \kw\ channels, horisontal red dashed lines indicate increase in GOES 1--8 \AA\ channel during $t_{90}$ , $\Delta$GOES, black dashed lines refer to peak interval in G2 channel, horisontal black dashed lines display HXR peak count rate determined as the sum of count rates in G1+G2+G3 channels, blue hatched area corresponds to HXR integral counts in G1+G2+G3 channels accumulated during $t_{90}$ determined by G2 channel. }
\end{figure}

\begin{figure*}\centering
\includegraphics[width=1.\textwidth]{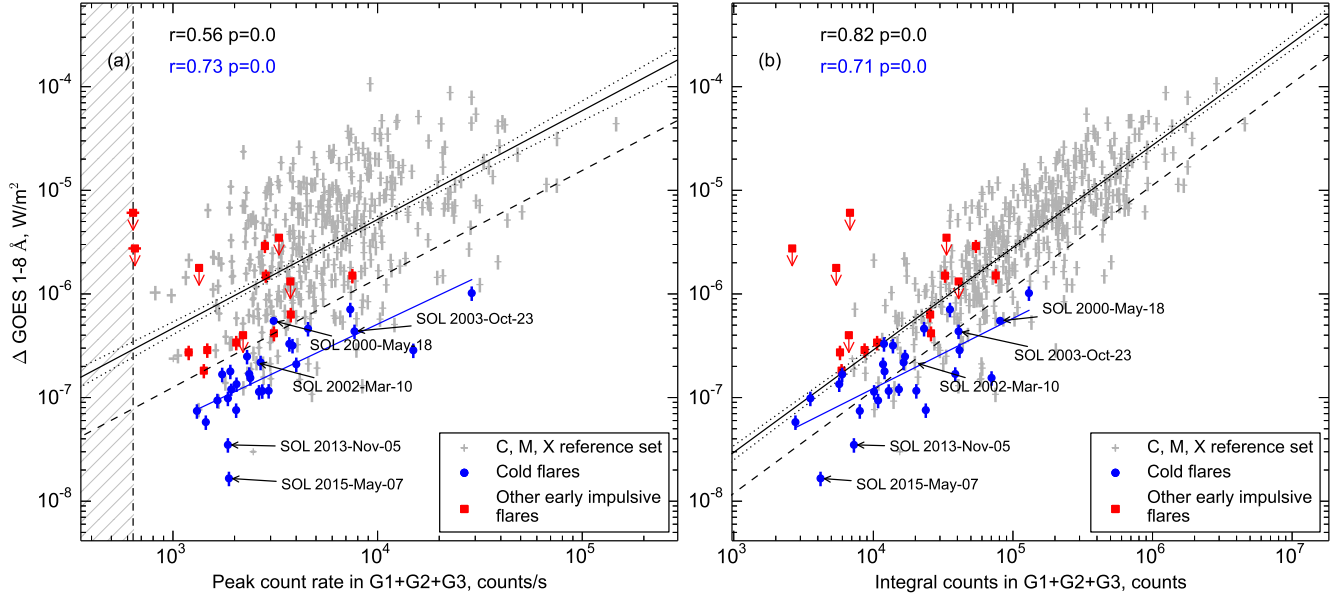}
\caption{\label{DeltaGOES_instr} Increase in GOES 1--8~\AA\ channel during impulsive phase of HXR emission vs. (a) HXR peak count rate, (b) HXR integral counts. Black and blue solid lines are linear dependences for all flares (both \eif s and flares from reference set) and CSFs only respectively, black dotted lines display 0.68 confidence band for all flares. Black and blue labels are correlation coefficients (r) and p-values (p) for all flares and for CSFs only correspondingly. Black dashed line represents the bound which separates `cold' outliers (see the text for more details).}
\end{figure*}

\section{\label{S_DAnalysis} Data Analysis}
\subsection{\label{HXR_vs_SXR} Relationships between HXR and SXR Emissions and Identification of the Cold Early Impulsive Flares}

To compare \kw\ and GOES time profiles, corrections for the light propagation time from the Wind spacecraft to the center of the Earth were applied to the \kw\ data. Such an approach gives an error within $\sim$20~ms for the light propagation to any ground-based or earth orbiting instrument, which is a satisfactory accuracy for the presented study.

The duration of the flare impulsive phase was determined in the \kw\ G2 channel, which boundaries changed from $\sim$50~keV--200~keV to $\sim$80~keV--300~keV during operational history. This channel was selected because it does not contain any contribution from the thermal emission.
The \kw\ background was approximated by a constant in time range in an interval selected within --1000~s and --200~s before the flare with fit probability $\geq$5~\%.  The duration of the flare impulsive phase was estimated using the so-called $t_{\rm 90}$, which is the difference between $t_{\rm 95}$, accumulation time of 95~\% of flare integral counts, and $t_{\rm 5}$, accumulation time of 5~\% of flare integral counts (see Fig~\ref{DeltaGOES_ex}). We employed $t_{\rm 90}$  because this value is less sensitive to the choice of signal-to-noise ratio than the total duration, $t_{\rm 100}$, \citep{Kouveliotou1993}. Values of  $t_{\rm 90}$ are listed in Table~\ref{table_sxr_hxr}.

To quantify the thermal response, which is needed to identify the outliers with a relatively low SXR  emission compared to the \eif s showing a more standard heating during the impulsive phase, we measured the increase of the GOES flux at the 1--8~\AA\ channel ($\Delta$GOES) during or following the impulsive HXR emission (see Figure~\ref{DeltaGOES_ex}).
In the case of constant background, the increase of GOES flux in 1--8 \AA\ channel was obtained as the difference between GOES flux at $t_{\rm 95}$ and flux at $t_{\rm 5}$. In the case of monotonically varying preflare GOES flux, the background was approximated by a 3-order polynomial and subtracted, and then the difference between fluxes at $t_{\rm 95}$ and $t_{\rm 5}$ was calculated. In the case of the absence of observable responce in the GOES 1--8 \AA\ channel  upper limits for $\Delta$GOES were estimated as 15~\% of GOES flux in 1--8 \AA\ channel which corresponds to GOES error in this channel \citep{Garcia1994}.

\begin{deluxetable*}{lcccccccc}[!t]
\tablecolumns{9}
\tablewidth{0pc}
\tablecaption{\label{table_sxr_hxr}Comparison of \csf properties in HXR and SXR ranges.
}
\tablehead{\colhead{N} & \colhead{Date}	& \colhead{$t_0$\tablenotemark{*}, hh:mm:ss} & \colhead{$t_{\rm 90}$} & \colhead{$\Delta$GOES} & \colhead{HXR peak} & \colhead{HXR int.} & \colhead{HXR peak\tablenotemark{**}} & \colhead{HXR int.\tablenotemark{**}}  \\
\colhead{} & \colhead{}	& \colhead{} & \colhead{(s)} & \colhead{10$^{-6}$~Wt/m$^{2}$} & \colhead{10$^3$~counts/s} & \colhead{10$^3$~counts} & \colhead{10$^{-6}$~erg/(s~cm$^2$)} & \colhead{10$^{-6}$~erg/(cm$^2$)}
}
\startdata
1 &	1998-May-07 & 05:32:37.072 & 3.8$\pm$0.5 & 0.33$\pm$0.05 & 3.70$\pm$0.08 & 11.92$\pm$0.15 & 6.36$\pm$0.05 & \nodata \\
2 &	1999-Jun-19 & 22:54:49.788 & 12.6$\pm$1.8 & 0.25$\pm$0.04 & 2.30$\pm$0.06 & 16.87$\pm$0.20  & \nodata & 18$\pm$3 \\
3 &	1999-Jul-30 & 22:58:09.675 & 2.3$\pm$0.3 & 0.098$\pm$0.016 & 1.86$\pm$0.06 & 3.54$\pm$0.09  & 3.08$\pm$0.07 & 5.9$\pm$0.9 \\
4 &	1999-Nov-09 & 08:26:21.703 & 8.1$\pm$0.4 & 0.71$\pm$0.11 & 7.35$\pm$0.09 & 35.39$\pm$0.22  & \nodata & \nodata \\
5 & 1999-Nov-14 & 14:55:08.244 & 3.0$\pm$0.3 & 0.058$\pm$0.009 & 1.45$\pm$0.06 & 2.77$\pm$ 0.09  & 2.41$\pm$0.07 & 4.6$\pm$0.6 \\
6 &	1999-Dec-02 & 20:01:00.012 & 3.4$\pm$0.8 & 0.135$\pm$0.022 & 2.04$\pm$0.06 & 5.70$\pm$0.11   & 3.39$\pm$0.07 & 6.1$\pm$1.7\\
7 &	2000-Mar-10 & 04:21:48.688 & 5.1$\pm$0.8 & 0.167$\pm$0.027 & 1.74$\pm$0.06 & 5.98$\pm$0.12  & 2.88$\pm$0.07 & 10.1$\pm$1.8\\
8 &	2000-Mar-18 & 02:25:10.567 & 23.6$\pm$2.5 & 0.169$\pm$0.027 & 2.36$\pm$0.06 & 38.62$\pm$0.28   & 3.94$\pm$0.10 & 57$\pm$10 \\
9 &	2000-May-18 & 07:21:59.706 & 14.9$\pm$2.5 & 0.0757$\pm$0.012 & 2.04$\pm$0.06 & 23.75$\pm$0.23  & 3.37$\pm$0.08 & 12$\pm$6 \\
10 & 2000-May-18 & 22:59:39.777 & 86$\pm$9 & 0.55\tablenotemark{***} & 3.11$\pm$0.07 & 81.0$\pm$0.5  & 5.26$\pm$0.15 & 357$\pm$55\\
11 & 2001-Oct-12 & 07:40:31.941 & 8.4$\pm$1.1 & 0.116$\pm$0.019 & 2.73$\pm$0.07 & 12.88$\pm$0.17  & 4.65$\pm$0.10 & 22$\pm$3 \\
12 & 2001-Nov-01 & 15:17:42.772 & 9.9$\pm$0.5 & 1.02$\pm$0.16 & 28.83$\pm$0.17 & 130.2$\pm$0.4  & \nodata & \nodata\\
13 & 2002-May-29 & 07:39:46.864 & 6.3$\pm$1.5 & 0.21$\pm$0.03 & 4.00$\pm$0.07 & 11.73$\pm$0.15  & 6.82$\pm$0.11 & 13$\pm$4 \\
14 & 2002-Aug-10 & 23:50:09.293 & 9.7$\pm$2.2 & 0.094$\pm$0.015 & 1.65$\pm$0.06 & 10.81$\pm$0.17  & \nodata & \nodata \\
15 & 2002-Aug-18 & 23:11:19.740 & 8.2$\pm$1.7 & 0.179$\pm$0.029 & 1.91$\pm$0.07 & 12.03$\pm$ 0.19  & \nodata & \nodata\\
16 & 2002-Aug-20 & 19:55:28.476 & 6.0$\pm$2.1 & 0.32$\pm$0.05 & 3.84$\pm$0.08 & 13.82$\pm$0.18  & \nodata & 23$\pm$9 \\
17 & 2003-Oct-23 & 22:17:39.620 & 11.6$\pm$2.4 & 0.44$\pm$0.07 & 7.71$\pm$0.10 & 40.78$\pm$ 0.25   & \nodata & 47$\pm$12\\
18 & 2005-Sep-08 & 02:15:49.996 & 8.3$\pm$2.6 & 0.29$\pm$0.05 & 14.94$\pm$0.14 & 41.44$\pm$0.29  & \nodata & 72$\pm$22\\
19 & 2011-Sep-19 & 07:43:40.791 & 14.9$\pm$2.8 & 0.12$\pm$0.02 & 1.93$\pm$0.06 & 15.24$\pm$0.20  & 3.26$\pm$0.07 & 29$\pm$7\\
20 & 2012-Jul-08 & 02:43:50.647 & 10.4$\pm$2.2 & 0.116$\pm$0.019 & 2.94$\pm$0.07 & 20.32$\pm$0.20  & 5.05$\pm$0.10 & 30$\pm$7\\
21 & 2013-Nov-05 & 03:50:24.588 & 6.12$\pm$1.6 & 0.035$\pm$0.006 & 1.86$\pm$0.06 & 7.26$\pm$0.13  & 3.22$\pm$0.04 & 5.0$\pm$1.5\\
22 & 2014-Jan-02 & 05:45:01.390 & 10.8$\pm$2.3 & 0.074$\pm$0.012 & 1.31$\pm$0.05 & 7.00$\pm$0.16  & 2.23$\pm$0.06 & 13$\pm$5 \\
23 & 2014-Jan-31 & 16:52:37.461 & 6.7$\pm$1.1 & 0.46$\pm$0.07 & 4.57$\pm$0.08 & 23.08$\pm$0.18   & 8.33$\pm$0.29 & 35$\pm$8\\
24 & 2014-Feb-08 & 05:49:29.848 & 7.2$\pm$2.2 & 0.114$\pm$0.018 & 2.63$\pm$0.06 & 10.12$\pm$0.15  & 4.48$\pm$0.09 & 6$\pm$3\\
25 & 2014-Oct-18 & 02:49:17.710 & 64$\pm$10 & 0.155$\pm$0.025 & 2.39$\pm$0.04 & 70.3$\pm$0.4  & \nodata & \nodata \\
26 & 2014-Oct-27 & 03:14:46.862 & 8.7$\pm$2.3 & 0.22$\pm$0.03 & 2.68$\pm$0.06 & 16.47$\pm$0.17   & 4.64$\pm$0.15 & 35$\pm$11\\
27 & 2015-May-07 & 12:41:40.415 & 3.3$\pm$0.7 & 0.0166$\pm$0.0027 & 1.88$\pm$0.06 & 4.18$\pm$0.10  & 3.27$\pm$0.04 & 7.2$\pm$1.7\\
\enddata

\tablenotetext{*}{The \kw\ trigger time after corrections for the light propagation to the Earth are applied.}
\tablenotetext{**}{Obtained using 3-channel fitting.}
\tablenotetext{***}{Upper limits.}

\end{deluxetable*}

\begin{figure*}[!b]\centering
\includegraphics[width=0.45\textwidth]{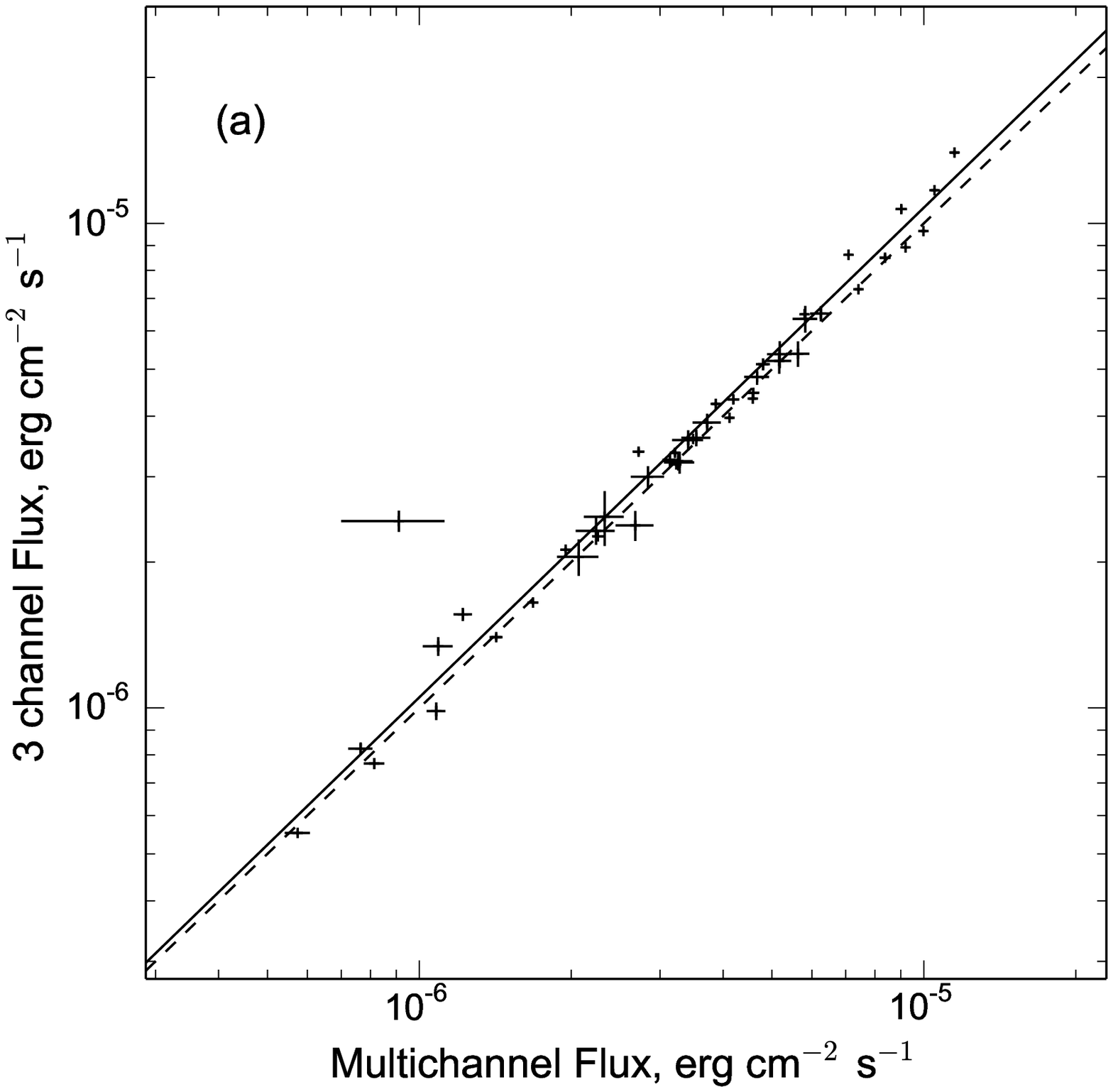}
\includegraphics[width=0.45\textwidth]{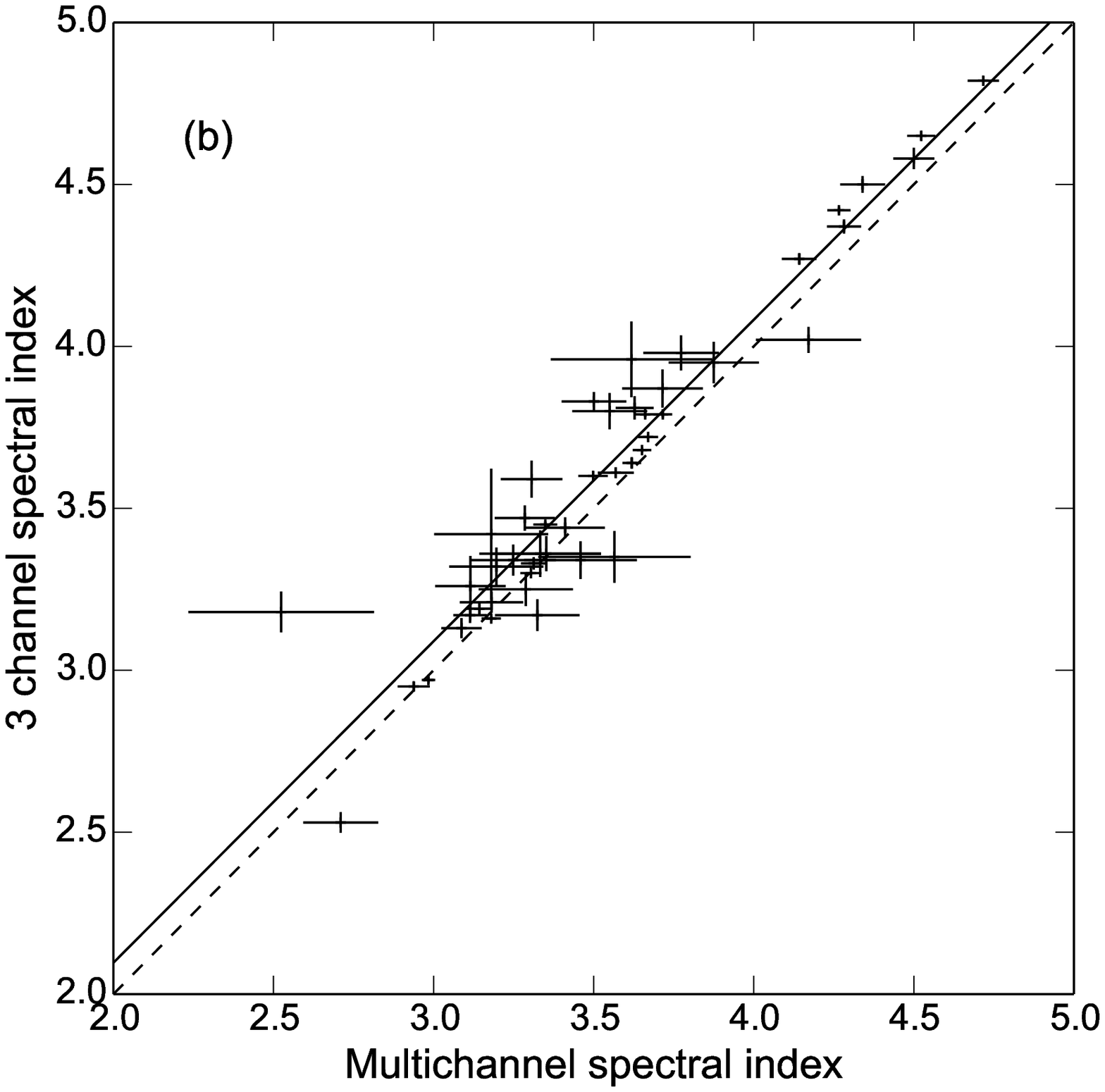}
\caption{\label{Mult_vs_3chan} Comparison between results of multichannel fitting and 3-channel fitting with power law model (a) for fluxes, (b) for spectral indices. Fitting was held on time frames corresponding to the accumulation time of peak multichannel spectra. Results with fit probability less than 0.01 were neglected. The dashed line reflects the equality and the solid line indicates linear regression between 3-channel and multichannel results.}
\end{figure*}

\begin{figure*}[!t]\centering
\includegraphics[width=1\textwidth]{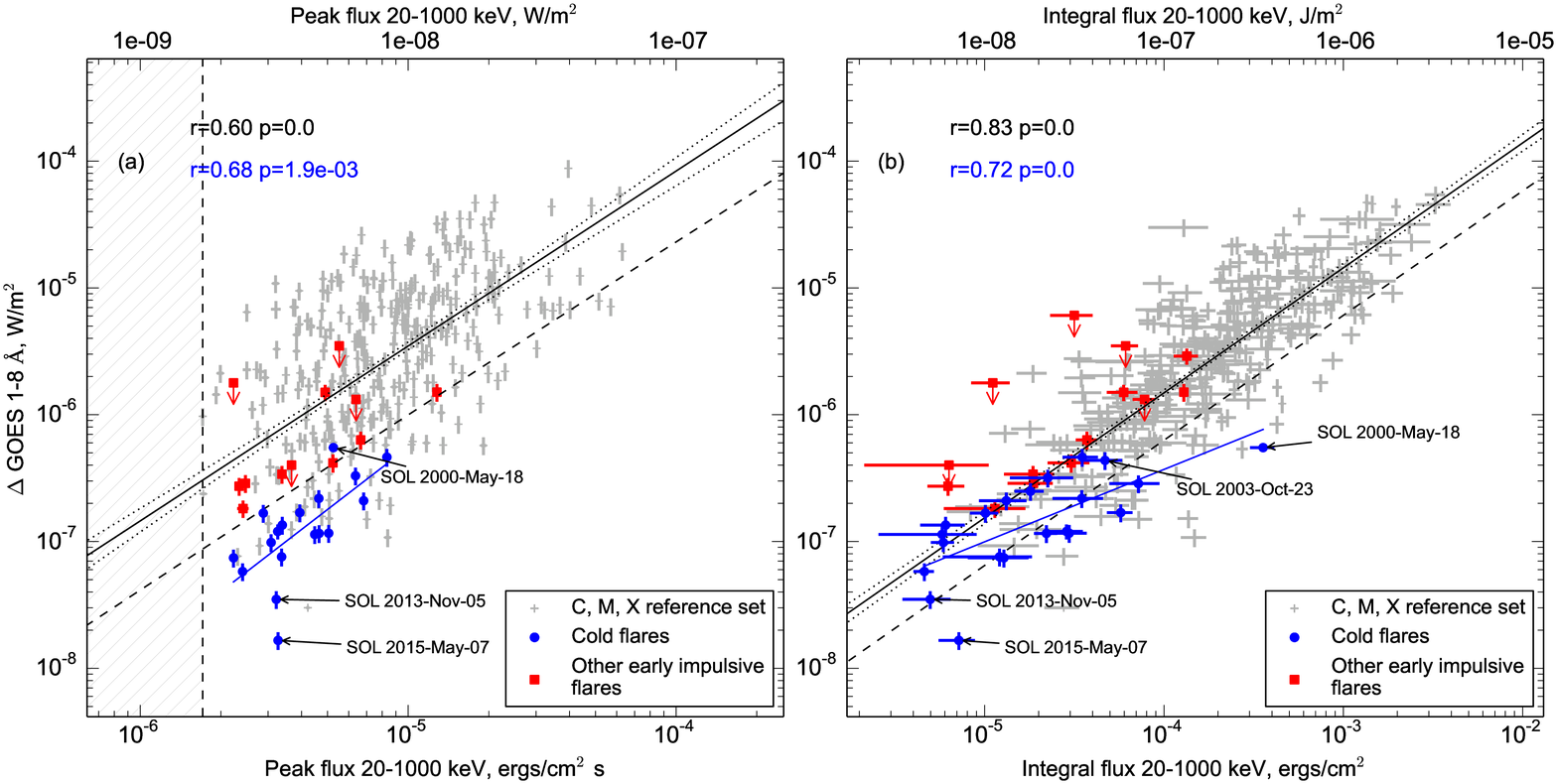}
\caption{\label{DeltaGOES_ufold} Increase in GOES 1--8~\AA\ channel during impulsive phase of HXR emission vs. (a) HXR peak flux, (b) HXR integral flux. Black and  blue solid lines are linear dependences for all flares (both \eif s and flares from reference set) and CSFs only respectively, black dotted lines display 0.68 confidence band for all flares. Black and blue labels are correlation coefficients (r) and p-values (p) for all flares and for CSFs only corresponingly. Black dashed line represents the bound which separates `cold' outliers (see the text for more details).}
\end{figure*}

The peak time frame was calculated on 1024 ms timescale according to counts in G2 channel, and the HXR peak count rate was taken as the sum of count rates in G1+G2+G3 channels after background substraction (see Figure~\ref{DeltaGOES_ex}). For one flare, the peak time frame was defined on 2.944~s timescale because of the failure in G2 trigger time history.

Results for $\Delta$GOES vs. HXR peak count rate regression for the \eif s and the reference set are shown in Figure~\ref{DeltaGOES_instr}(a) and are listed in Table~\ref{table_sxr_hxr}. The grey hatched area in the left of the figure indicates the feature, that the Konus-Wind cannot register flares with that low HXR peak count rates in the triggered mode. We calculated regression coefficients and confidence interval using python \verb"bces" \citep{Akritas1996, Nemmen2012} procedure, which takes into account both x and y uncertanties, and paired-bootstrap resampling. The black solid line in the Figure corresponds to the linear regression between $\Delta$GOES and HXR peak count rate for all flares and dotted lines indicate 68~\% confidence band. 
The flares lying near the regression line can be referred as ones with `mean' thermal response, while those far above the line can be interpreted as thermal dominated, and the flares far below regression line are outliers, which demonstrate lack of thermal emission. 
The quantitative criterion to select these outliers was as follows.
We built distribution of distances between the regression line and each flare with negative distances for the flares below regression line. 
We considered 68~\% of flares, those which lie between 16 and 84\,\% quantile of distances distribution, have average thermal response, while \eif s below 16~\% quantile (dashed line on the Figure) are nonthermal dominated, i. e. `cold' \eif s.

The blue solid line in the Figure is a rough estimation of the linear regression between $\Delta$GOES and HXR peak count rate for cold flares only. Several events are individually labeled in this plot: these are flares 2013-Nov-05 and 2015-May-07, which demonstrate very low thermal response, the flare 2002-Mar-10 from \cite{Fl_etal_2016}, and flares 2000-May-18, 22:59 UT, and 2003-Oct-23 that will be discussed later. Labels ``$r$'' and ``$p$'' denote the Pearson correlation coefficient and p-values (probabilities that correlation between parameters is elusive) for all events (black) and for cold flares only (blue). Correlation coefficients were calculated using python \verb"scipy.stats.linregress" function between decimal logarithms of observed values hereafter. Correlation coefficient for the cold flares, $r$=0.73, is larger than coefficient for all events is $r$=0.56, p-values are negligible in both cases.

The \kw\ HXR integral counts during the impulsive flare phase were estimated as the  background subtracted sum of G1+G2+G3 counts during $t_{\rm 90}$ defined for G2 channel. Relationship between the HXR integral counts and $\Delta$GOES is presented in Figure~\ref{DeltaGOES_instr}(b) and Table~\ref{table_sxr_hxr}. As in Figure~\ref{DeltaGOES_instr}(a), the black and blue solid lines represent linear regressions for all flares and the cold flares, respectively; dotted lines represent 68~\% confidence band for all flares, which were obtained similarly to Figure~\ref{DeltaGOES_instr}(a). Meanings of the labels are the same as those in Figure~\ref{DeltaGOES_instr}(a).  The largest correlation coefficient, $r$=0.82, is for all flares, $r$=0.71 for CSFs, p-values for these two groups are close to zero. Outliers for this relationship were defined in the same manner as for $\Delta$GOES vs. HXR peak count rate relationship. For the most cases, the `cold' early impulsive outliers in Figure~\ref{DeltaGOES_instr}(b) are also the outliers in panel (a). The only exception is the event of 2000-May-18, 22:59 UT.

As the \kw\ energy boundaries changed with time it is reasonable  to compare the relationships obtained from the instrumental HXR characteristics, such as  HXR peak count rate and HXR integral counts, with those obtained from unfolded characteristics, namely HXR peak flux and HXR integral flux. To find HXR fluxes in physical units (for example, erg  ~s$^{\rm -1}$~cm$^{\rm -2}$) especially in the case of rather broad energy channels, a spectral model should be selected, then spectral fitting performed, and HXR fluxes calculated based on the obtained fitting parameters. To keep uniformity of the fitting results, 3-channel fitting rather than multichannel one was employed, because a fraction of the flare impulsive phase might have occurred  before the \kw\ trigger, i.e. in the waiting mode data, where no multichannel spectra are available.

Results of 3-channel and multichannel fitting with power-law model  were compared for peak spectra of \eif s and some reference flares (see Section~\ref{HXR_fit}). Results with fit probabilities $<$1~\% were neglected. Comparisons between  the results of the 3-channel and multichannel fitting are presented in Figure~\ref{Mult_vs_3chan}; namely, comparison between the power-law indices (left) and between HXR fluxes in 20--1000~keV range (right). Errors on both plots refer to 68~\% confidence level. Dashed line indicates the expected equality of fit results, while the solid line refers to the linear regression between multichannel and 3-channel fit results. As can be seen from  Figure~\ref{Mult_vs_3chan}, the results of multichannel and 3-channel fitting are in good agreement with the exception of one event. Thus, we do not apply any extra corrections to 3-channel fitting results.

Relationships between $\Delta$GOES and the HXR peak flux or the integral flux are presented in Figure~\ref{DeltaGOES_ufold} and Table~\ref{table_sxr_hxr}. Meanings of labels and lines are the same as in Figure~\ref{DeltaGOES_instr}(a).
From comparison between Figure~\ref{DeltaGOES_instr}(a) and Figure~\ref{DeltaGOES_ufold}(a) it is clear that the same flares form the set of `cold flare' outliers in both plots. Thus, the  \kw\ energy boundary variations do not significantly affect our selection of the CSFs. The values of the correlation coefficients for regressions shown in Figures~\ref{DeltaGOES_instr}(a) and \ref{DeltaGOES_ufold}(a) are also close to each other; specifically, for all flares: $r$=0.60 for the HXR peak flux vs $r$=0.56 for the HXR peak count rate, p-value is also negligible and for the  CSFs: $r$=0.68 vs $r$=0.73 and p-value is 1.9e-3.

The situation is similar comparing Figures~\ref{DeltaGOES_instr}(b) and Figures~\ref{DeltaGOES_ufold}(b): the correlation coefficient for all flares remains $r$=0.83 and the CSFs correlation coefficient also didn't change significantly and probabilities, that correlations are elusive, are also close to zore for all three groups. But some CSFs move closer to the main regression line comparing to relation between $\Delta$GOES and HXR integral counts. This can be caused by underestimation of integral flux while averaging flux over flare duration in the fitting procedure.

Thus for flares, for which unfolded HXR fluxes were obtained, we conclude that the \kw\ energy boundaries evolution does not affect the proportions between HXR and SXR characteristics considerably. So, for selection of CSF group we formulate the criterion  based on instrumental HXR estimations: at least one of the following statements must hold for a flare to be classified as ``cold''---either (1) the ratio between $\Delta$GOES and HXR peak flux or (2) the ratio between $\Delta$GOES and HXR integral counts is lower than that for the majority (84~\%) of flares.
This approach is more universal than using the unfolded data, because these observational measures can be obtained for all events. This approach yielded 27 CSFs listed in Table~\ref{table_KW_flare_list}, on which we focus in this paper.

\begin{figure}[!t]\centering
\includegraphics[width=0.5\textwidth]{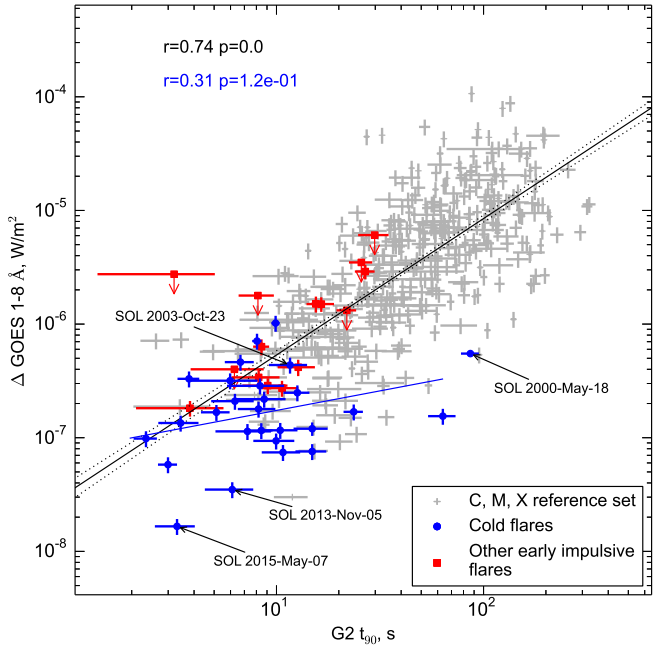}
\caption{\label{DeltaGOES_t90}  Increase in GOES 1--8~\AA\ channel during impulsive phase of HXR emission vs. $t_{90}$. Black and blue solid lines are linear dependences for all flares (both \eif s and flares from reference set) and cold flares respectively, black dotted lines display 0.68 confidence band for all flares. Black and blue labels are correlation coefficients (r) and p-values (p) for all flares and for cold flares corresponingly.}
\end{figure}

Relationship between increase in GOES channel 1--8 \AA\ and $t_{90}$ is plotted in Figure~\ref{DeltaGOES_t90}. As expected, a high correlation between $\Delta$GOES and $t_{90}$ is observed for all flares $r=0.74$. CSFs are grouped in the area of low $\Delta$GOES and low $t_{90}$ excepting two events and thus do not demonstrate significant correlation, $r=0.31$.

\subsection{\label{HXR_fit} Spectral properties of the  HXR bursts}

The HXR spectral analysis was perfomed using the \kw\ multichannel data for 25 of 27 CSFs, because for 2 event spectral data were damaged. In addition, multichannel fits were obtained for 71 events from non-cold-flare reference set occurred during years 2010--2016.

The HXR spectrum fitting was performed on the peak spectra;  the  peaks were defined according to the  count rates in G2 channel (see Section~\ref{HXR_vs_SXR}). The photon energy range between 20 and 1000~keV was considered for analysis of both cold and reference flares: 20~keV corresponds to the low-energy boundary of  \kw, while no photons above 1000~keV were considered, even when present, because the nuclear deexcitation line emission may contribute to the spectrum at those high energies in addition to the nonthermal electron bremsstrahlung. The spectral channels were grouped to have a minimum of 10 counts per channel to ensure the validity of the $\chi^2$ statistic.
The spectral analysis was performed using \verb"XSPEC 12.9.0" \citep{Arnaud1996}.

\begin{figure}[!b]
\includegraphics[scale=0.45]{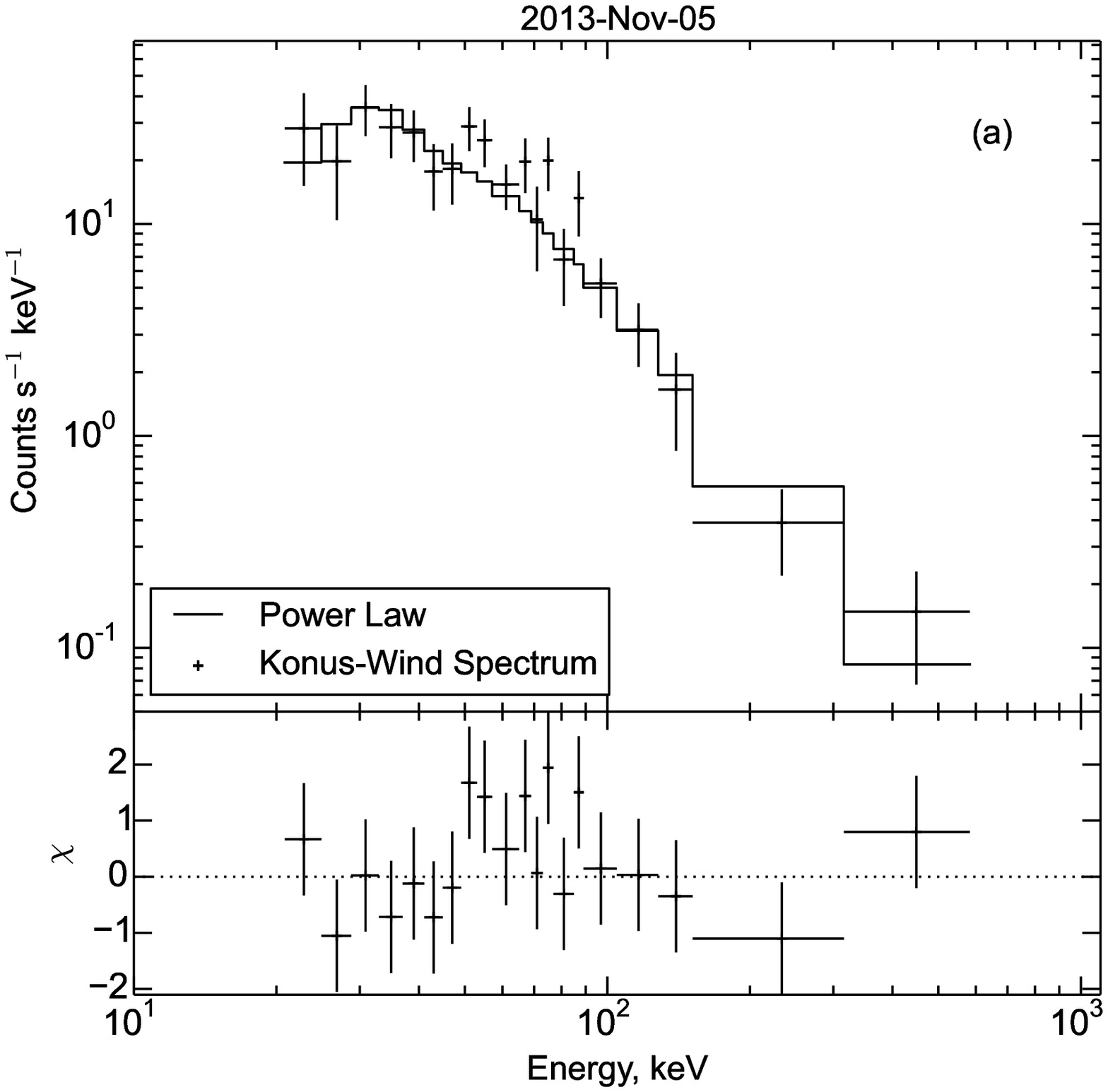}
\includegraphics[scale=0.45]{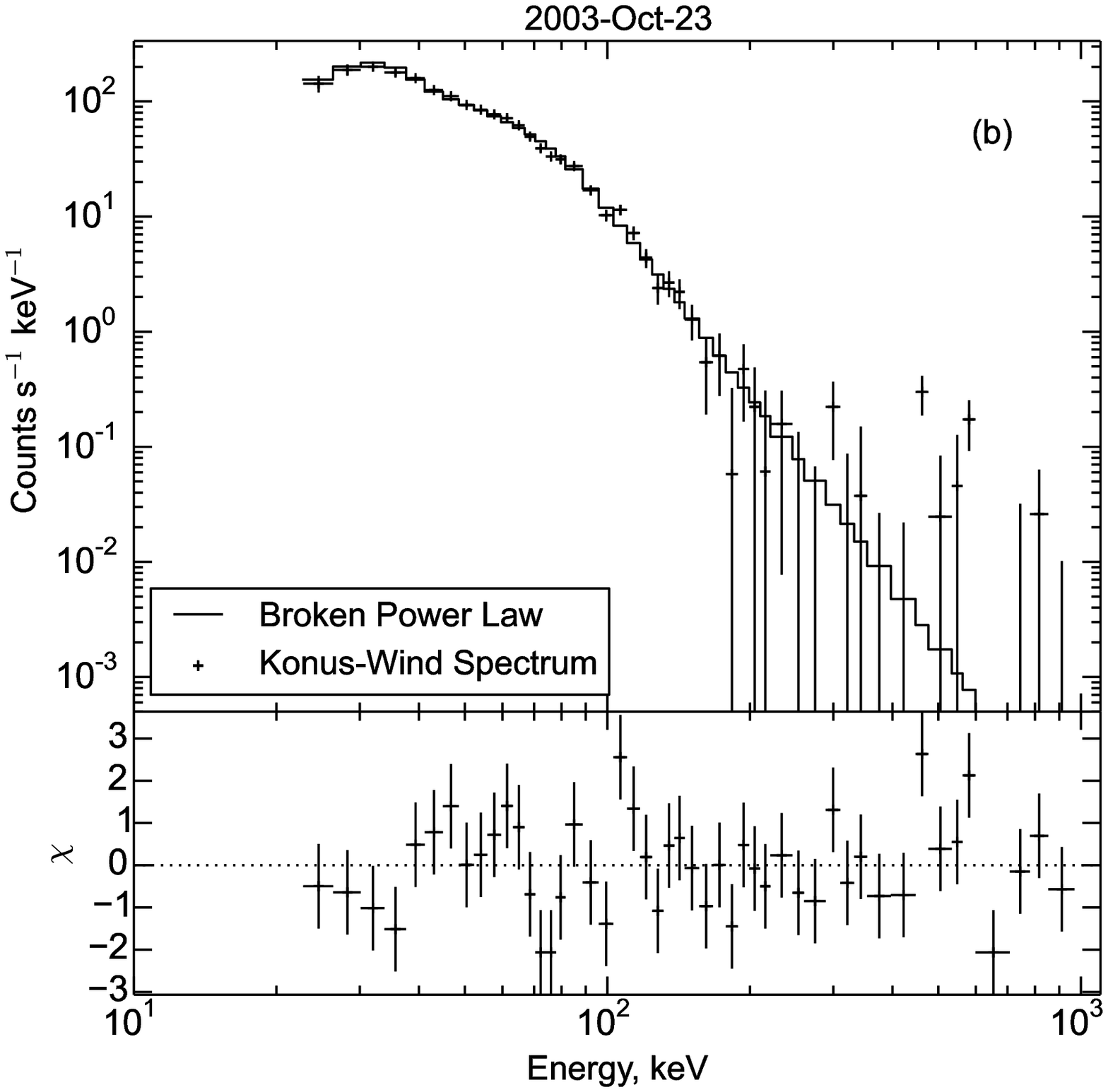}
\caption{\label{fig_HXR_example_phot} Examples of HXR CSFs spectra fittes by (a) PL and (b) 2PL models.
}
\end{figure}


\begin{deluxetable*}{lcccccccc}[!t]
\tablecolumns{9}
\tablewidth{0pc}
\tablecaption{\label{table_hxr_pl}PL and 2PL fit results.
}
\tablehead{\colhead{N} & \colhead{Date}	& \colhead{$t_0$\tablenotemark{*}, hh:mm:ss} & \colhead{$\gamma_1$} & \colhead{$E_{\rm break,ph}$} & \colhead{$\gamma_2$} & \colhead{Photon flux} & \colhead{$\chi^2$/dof} & \colhead{Prob.}  \\
\colhead{} & \colhead{}	& \colhead{} & \colhead{} & \colhead{keV} & \colhead{} & \colhead{10$^{-6}$~erg/(s~cm$^2$)} & \colhead{} & \colhead{}
}
\startdata
1 &	1998-May-07 & 05:32:37.072 & 2.51$_{-0.12}^{+0.11}$ & 67$_{-6}^{+6}$ & 3.67$_{-0.18}^{+0.22}$ & 4.76$_{-0.16}^{+0.16}$ & 52.53/57 & 6.4e-01  \\
2 &	1999-Jun-19 & 22:54:49.788 & 2.68$_{-0.08}^{+0.09}$ & 101$_{-11}^{+19}$ & 4.5$_{-0.4}^{+0.9}$ & 2.76$_{-0.07}^{+0.07}$ & 57.10/57 & 4.7e-01  \\
3 &	1999-Jul-30 & 22:58:09.675 & 3.26$_{-0.15}^{+0.17}$ & \nodata & \nodata & 2.81$_{-0.21}^{+0.21}$ & 11.27/26 & 9.9e-01  \\
4 &	1999-Nov-09 & 08:26:21.703 & 2.49$_{-0.13}^{+0.13}$ & 66$_{-6}^{+8}$ & 3.86$_{-0.12}^{+0.18}$ & 10.75$_{-0.27}^{+0.28}$ & 56.57/41 & 5.3e-02 \\
5 & 1999-Nov-14 & 14:55:08.244 & 2.52$_{-0.23}^{+0.29}$ & \nodata & \nodata & 0.91$_{-0.21}^{+0.21}$ & 24.12/26 & 5.7e-01 \\
6 &	1999-Dec-02 & 20:01:00.012 & 3.46$_{-0.16}^{+0.18}$ & \nodata & \nodata & 2.84$_{-0.21}^{+0.21}$ & 23.47/25 & 5.5e-01 \\
7 &	2000-Mar-10 & 04:21:48.688 & 3.29$_{-0.14}^{+0.15}$ & \nodata & \nodata & 3.26$_{-0.22}^{+0.22}$ & 28.00/26 & 3.6e-01 \\
8 &	2000-Mar-18 & 02:25:10.567 & 3.17$_{-0.14}^{+0.22}$ & 60$_{-5}^{+16}$ & 4.7$_{-0.2}^{+0.9}$ & 2.88$_{-0.11}^{+0.17}$ & 50.41/57 & 7.2e-01 \\
9 &	2000-May-18 & 07:21:59.706 & 3.13$_{-0.27}^{+0.09}$ & 72$_{-15}^{+6}$ & 5.1$_{-0.7}^{+0.5}$ & 2.74$_{-0.15}^{+0.09}$ & 78.38/58 & 3.9e-02 \\
10 & 2000-May-18 & 22:59:39.777 & 3.77$_{-0.11}^{+0.12}$ & \nodata & \nodata & 5.6$_{-0.3}^{+0.3}$ & 30.34/25 & 2.1e-01 \\
11 & 2001-Oct-12 & 07:40:31.941 & 3.12$_{-0.10}^{+0.11}$ & \nodata & \nodata & 4.67$_{-0.26}^{+0.26}$ & 35.23/27 & 1.3e-01 \\
12 & 2001-Nov-01 & 15:17:42.772 & \nodata & \nodata & \nodata & \nodata &  \nodata & \nodata \\
13 & 2002-May-29 & 07:39:46.864 & 3.29$_{-0.09}^{+0.09}$ & \nodata & \nodata & 6.3$_{-0.3}^{+0.3}$ & 29.83/25 & 2.3e-01 \\
14 & 2002-Aug-10 & 23:50:09.293 & \nodata & \nodata & \nodata & \nodata &  \nodata & \nodata \\
15 & 2002-Aug-18 & 23:11:19.740 & 3.17$_{-0.17}^{+0.18}$ & \nodata & \nodata & 2.66$_{-0.25}^{+0.25}$ & 31.69/26 & 2.0e-01 \\
16 & 2002-Aug-20 & 19:55:28.476 & 2.30$_{-0.26}^{+0.24}$ & 64$_{-8}^{+12}$ & 4.2$_{-0.4}^{+0.7}$ & 6.2$_{-0.4}^{+0.4}$ & 45.69/31 & 4.3e-02 \\
17 & 2003-Oct-23 & 22:17:39.620 & 2.94$_{-0.08}^{+0.07}$ & 74$_{-5}^{+4}$ & 5.12$_{-0.27}^{+0.28}$ & 11.53$_{-0.27}^{+0.28}$ & 57.63/44 & 8.2e-02 \\
18 & 2005-Sep-08 & 02:15:49.996 & 2.08$_{-0.11}^{+0.07}$ & 106$_{-20}^{+15}$ & 3.27$_{-0.24}^{+0.26}$ & 16.9$_{-0.5}^{+0.5}$ & 80.94/53 & 8.0e-03 \\
19 & 2011-Sep-19 & 07:43:40.791 & 3.20$_{-0.14}^{+0.15}$ & \nodata & \nodata & 3.28$_{-0.22}^{+0.22}$ & 18.17/23 & 7.5e-01 \\
20 & 2012-Jul-08 & 02:43:50.647 & 3.31$_{-0.13}^{+0.14}$ & \nodata & \nodata & 3.81$_{-0.25}^{+0.25}$ & 47.16/25 & 4.7e-03 \\
21 & 2013-Nov-05 & 03:50:24.588 & 2.63$_{-0.12}^{+0.13}$ & \nodata & \nodata & 2.87$_{-0.23}^{+0.23}$ & 26.50/24 & 3.3e-01 \\
22 & 2014-Jan-02 & 05:45:01.390 & 3.18$_{-0.16}^{+0.18}$ & \nodata & \nodata & 2.07$_{-0.20}^{+0.20}$ & 32.27/22 & 7.3e-02 \\
23 & 2014-Jan-31 & 16:52:37.461 & 4.17$_{-0.11}^{+0.11}$ & \nodata & \nodata & 7.8$_{-0.4}^{+0.4}$ & 44.75/21 & 1.9e-03 \\
24 & 2014-Feb-08 & 05:49:29.848 & 3.32$_{-0.12}^{+0.13}$ & \nodata & \nodata & 3.54$_{-0.23}^{+0.23}$ &  36.4/24 & 5.0e-02 \\
25 & 2014-Oct-18 & 02:49:17.710 & \nodata & \nodata & \nodata & \nodata &  \nodata & \nodata \\
26 & 2014-Oct-27 & 03:14:46.862 & 3.32$_{-0.22}^{+0.24}$ & \nodata & \nodata & 2.14$_{-0.26}^{+0.26}$ & 16.99/22 & 7.6e-01 \\
27 & 2015-May-07 & 12:41:40.415 & 2.73$_{-0.14}^{+0.15}$ & \nodata & \nodata & 2.61$_{-0.24}^{+0.24}$ & 29.40/27 & 3.4e-01 \\
\enddata

\tablenotetext{*}{The \kw\ trigger time after corrections for the light propagation to the Earth are applied.}

\end{deluxetable*}

Initially, we attempted fitting the spectra using the single  power-law (PL) model, but the peak spectra were inconsistent with this simple model for many flares because of a  spectral break. Thus, for spectra inconsistent with PL we used the  phenomenological broken power-law model, 2PL:

\begin{equation}
\label{eq_2PL}
I(E) = \begin{cases} A \left(\frac{E}{100keV}\right)^{-\gamma_1} & E \le E_{break, ph} \\
 A E_{break, ph}^{\gamma_2-\gamma_1} \left(\frac{E}{100keV}\right)^{-\gamma_2} & E_{break, ph} < E \end{cases}
\end{equation}
where $A$ is the normalization at 100 keV in units of photons~cm$^{-2}$~s$^{-1}$~keV$^{-1}$.
Photon flux was calculated in 20--1000~keV range using \verb"cflux" convolution model in XSPEC assuming spectrum integration within 20 and 1000~keV.

In cases when both PL and 2PL models were consistent with the data, the preferred model was chosen according to $F$-test \citep{Bevington1969}, i.e. the decrease in $\chi^2$ vs. the degrees of freedom decrease: the criterion for accepting 2PL model (a model with 2 additional free  parameters compared to the PL model) was a decrease of $\chi^2>13.5$ using 2PL model compared to using PL model which corresponds to chance probability of such decrease p$\leq$0.1~\%. One CSFs and 3 flares from the reference group with fit probabilities for both PL and 2PL model p$\leq$0.1~\% were excluded.

\begin{figure*}[!t]\centering
\includegraphics[width=1\textwidth]{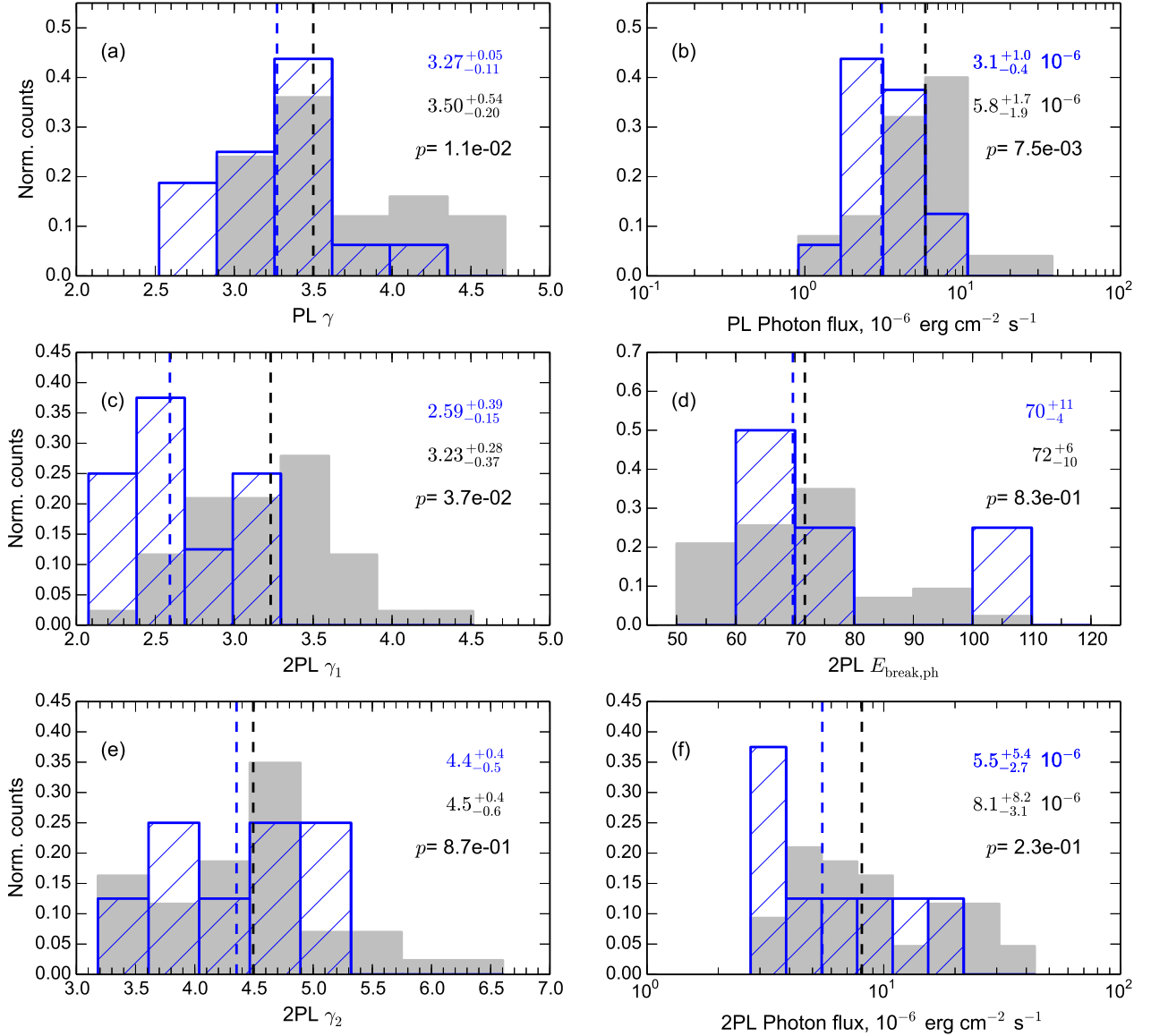}
\caption{\label{HXR_dist_phot} Parameter distributions for HXR photon spectra. Bin heights are normalized to total number of events in each group. Blue hatched histograms refer to cold flares, grey histograms -- to reference group. Median values and 0.5 probability ranges are presented on each plot for cold flares and for reference group in blue and black letters respectively. Black "p"\ indicates two-sided p-value from Kolmogorov-Smirnov test in suggestion that distributions of a given parameter for cold flares and reference flares are equal. (a) PL photon spectral index, $\gamma$, distributions; (b) PL photon flux distributions in 20--1000~keV range; (c) 2PL photon spectral index in lower energy range, $\gamma_1$, distributions; (d) 2PL break energy of photon spectrum, E$_{break,ph}$, distributions; (e) 2PL photon spectral index in higher energy range, $\gamma_2$, distributions; (f) 2PL photon flux distributions in 20--1000~keV range.
}
\end{figure*}

Examples of CSFs spectra fitted by PL and 2PL models are presented in Figure~\ref{fig_HXR_example_phot} and the fit results for all CSFs with successful fits are listed in Table~\ref{table_hxr_pl}. The PL model successfully  fits 16 of 25 CSFs and 25 of 71 reference flares. Distributions of obtained fitting parameters $\gamma$ and photon flux are presented in Figure~\ref{HXR_dist_phot}(a--b). On each plot, the  median values and 50~\% ranges are marked for cold flares (blue) and reference flares (grey), the value ``$p$'' denotes probability obtained by the Kolmogorov-Smirnov test hereafter in the assumption that distributions for cold flares and reference set are similar. The Kolmogorov-Smirnov test was performed using python \verb"scipy.stats.ks_2samp" function. The  median values of the photon spectral indices $\gamma$ for the  CSFs and reference flares are equal to  3.3 and 3.5, respectively; though the difference between the  median values is not high, but 50~\% range is narrower for CSFs and $\gamma$ distribution is shifted towards lower values, thus CSFs are significantly harder (probability that two distributions are equal is $\sim$1~\%). The median value of the photon flux for   CSFs is almost two times lower than for the reference set, 3.1$\times$10$^{-6}$ vs. 5.8$\times$10$^{-6}$~erg~cm$^2$~s$^{-1}$; there are no events with peak fluxes greater than 10.0$\times$10$^{-6}$~erg~cm$^2$~s$^{-1}$ among CSFs, while the peak fluxes for reference flares extend up to 40.0$\times$10$^{-6}$~erg~cm$^2$~s$^{-1}$, p-value is also low, 0.75~\%.

Fitting using the 2PL model was successful for 8 CSFs and 43 flares from the reference set. This means that the 2PL model was required much more rarely for CSFs than for the reference flares. This might happen, at least for a fraction of flares, because the CSFs  are lower intensity flares, for which low signal-to-noise ratio in higher energy range did not allow to detect the energy break even if existed.  Distributions of the fitting parameters $\gamma_1$, $E_{\rm break,ph}$, $\gamma_2$ and photon flux for 2PL are plotted in Figure~\ref{HXR_dist_phot}(c--f). Distributions of low energy photon power-law index $\gamma_1$ show that for 2PL model CSFs are also harder: the  median value for CSFs is 2.6 vs. 3.2 for the reference group and the overall shape of the  histogram is shifted towards lower values of $\gamma_1$ as compared to the reference flares, p-value is $p$=3.7~\%.  The median values of photon break energies $E_{\rm break,ph}$ for both groups are close to 70~keV. For CSFs there are no events with breaks below 60~keV and for the reference group there are no events with break below 50~keV. Probability of $E_{\rm break,ph}$ distributions coincidence for CSFs and reference set is 83~\%. The median value of the high-energy photon power-law index, $\gamma_2$, for CSFs is also a slightly  harder than for the reference set, 4.4 vs. 4.5, but this difference is insignificant, the probability that $\gamma_2$ distributions for CSFs and reference set distributions are equal, is very high:  87~\%. Photon flux distributions are presented in Figure~\ref{HXR_dist_phot}(f), these values may differ from those listed in Table~\ref{table_sxr_hxr}, because they were obtained on a different timescales. The median value for the reference set is 8.1$\times$10$^{-6}$~erg~cm$^{-2}$~s$^{-1}$  which is higher than for CSFs, 5.5$\times$10$^{-6}$~erg~cm$^{-2}$~s$^{-1}$, but also it could be insignificant, $p$=23~\%.
For both  CSFs and  reference flares $\gamma_2$ is always  larger (by absolute value) than $\gamma_1$; thus, no break-ups were observed in the entire analyzed data set. A likely reason for this finding, that is in apparent contrast with many reports of HXR spectra with break-ups, is the trigger nature of \kw\  that only records flares with reasonably hard spectra, while the break-ups are typically observed when the low-energy part of the HXR spectrum is steep.

\begin{deluxetable*}{lcccccccc}
\tablecolumns{9}
\tablewidth{0pc}
\tablecaption{\label{table_hxr_brm}brmPL and brm2PL fit results.
}
\tablehead{\colhead{N} & \colhead{Date}	& \colhead{$t_0$\tablenotemark{*}, hh:mm:ss}  & \colhead{$\delta_1$} & \colhead{$E_{\rm break,el}$} & \colhead{$\delta_2$} & \colhead{Electron flux} & \colhead{$\chi^2$/dof} & \colhead{Prob.} \\
\colhead{} & \colhead{}	& \colhead{}  & \colhead{} & \colhead{keV} & \colhead{} & \colhead{10$^{35}$~el/s} & \colhead{} & \colhead{}
}
\startdata
1 &	1998-May-07 & 05:32:37.072 & 2.7$_{-0.5}^{+0.3}$ & 188$_{-27}^{+37}$ & 5.3\tablenotemark{**} & 0.8$_{-0.5}^{+0.7}$ & 52.51/58 & 6.8e-01 \\
2 &	1999-Jun-19 & 22:54:49.788 & 2.7$_{-0.3}^{+0.2}$ & 242$_{-21}^{+25}$ & 6.6\tablenotemark{**} & 0.41$_{-0.16}^{+0.22}$ & 48.00/58 & 8.1e-01 \\
3 &	1999-Jul-30 & 22:58:09.675 & 4.09$_{-0.15}^{+0.16}$ & \nodata & \nodata & 6.5$_{-1.9}^{+2.7}$ & 11.18/26 & 9.9e-01 \\
4 &	1999-Nov-09 & 08:26:21.703 & 2.8$_{-0.3}^{+0.2}$ & 213$_{-44}^{+56}$ & 6.3$_{-0.9}^{+3.3}$ & 2.3$_{-1.0}^{+1.0}$ & 39.30/41 & 5.4e-01 \\
5 & 1999-Nov-14 & 14:55:08.244 & 3.45$_{-0.24}^{+0.24}$ & \nodata & \nodata & 0.39$_{-0.22}^{+0.39}$ & 24.44/26 & 5.5e-01 \\
6 &	1999-Dec-02 & 20:01:00.012 & 4.29$_{-0.15}^{+0.17}$ & \nodata & \nodata & 9.8$_{-2.8}^{+4.1}$ & 23.41/25 & 5.5e-01 \\
7 &	2000-Mar-10 & 04:21:48.688 & 4.12$_{-0.13}^{+0.14}$ & \nodata & \nodata & 8.0$_{-2.1}^{+2.8}$ & 27.83/26 & 3.7e-01 \\
8 &	2000-Mar-18 & 02:25:10.567 & 3.2$_{-0.4}^{+0.3}$ & 124$_{-10}^{+12}$ & 6.2\tablenotemark{**} & 2.0$_{1.0}^{+1.3}$ & 43.40/58 & 9.2e-01 \\
9 &	2000-May-18 & 07:21:59.706 & 3.0$_{-0.3}^{+0.3}$ & 145$_{-11}^{+12}$ & 7.1\tablenotemark{**} & 1.3$_{-0.5}^{+0.7}$ & 67.70/59 & 2.0e-01 \\
10 & 2000-May-18 & 22:59:39.777 & 4.59$_{-0.11}^{+0.12}$ & \nodata & \nodata & 35$_{-7}^{+9}$ & 29.73/25 & 2.3e-01 \\
11 & 2001-Oct-12 & 07:40:31.941 & 3.96$_{-0.10}^{+0.10}$ & \nodata & \nodata & 8.0$_{-1.7}^{+2.2}$ & 35.23/27 & 1.3e-01 \\
12 & 2001-Nov-01 & 15:17:42.772 & \nodata & \nodata & \nodata & \nodata &  \nodata & \nodata \\
13 & 2002-May-29 & 07:39:46.864 & 4.12$_{-0.09}^{+0.09}$ & \nodata & \nodata & 15.3$_{-2.8}^{+3.4}$ & 29.91/25 & 2.3e-01 \\
14 & 2002-Aug-10 & 23:50:09.293 & \nodata & \nodata & \nodata & \nodata &  \nodata & \nodata \\
15 & 2002-Aug-18 & 23:11:19.740 & 4.02$_{-0.16}^{+0.18}$ & \nodata & \nodata & 5.2$_{-1.7}^{+2.5}$ & 31.78/26 & 2.0e-01 \\
16 & 2002-Aug-20 & 19:55:28.476 & 3.77$_{-0.08}^{+0.08}$ & \nodata & \nodata & 8.1$_{-1.5}^{+1.9}$ & 45.06/33 & 7.9e-02 \\
17 & 2003-Oct-23 & 22:17:39.620 & 2.5$_{-0.5}^{+0.4}$ & 140$_{-16}^{+22}$ & 6.8$_{-0.7}^{+1.6}$ & 2.1$_{-1.0}^{+1.0}$ & 46.00/44 & 3.9e-01 \\
18 & 2005-Sep-08 & 02:15:49.996 & 2.0$_{-0.4}^{+0.3}$ & 390$_{-45}^{+53}$ & 5.8\tablenotemark{**} & 0.31$_{-0.15}^{+0.23}$ & 71.10/54 & 5.9e-02 \\
19 & 2011-Sep-19 & 07:43:40.791 & 4.03$_{-0.13}^{+0.14}$ & \nodata & \nodata & 6.6$_{-1.8}^{+2.4}$ & 18.16/23 & 7.5e-01 \\
20 & 2012-Jul-08 & 02:43:50.647 & 4.15$_{-0.19}^{+0.22}$ & \nodata & \nodata & 10$_{-4}^{+6}$ & 47.00/25 & 5.0e-03 \\
21 & 2013-Nov-05 & 03:50:24.588 & 3.50$_{-0.11}^{+0.12}$ & \nodata & \nodata & 1.5$_{-0.4}^{+0.6}$ & 26.50/24 & 3.3e-01 \\
22 & 2014-Jan-02 & 05:45:01.390 & 4.03$_{-0.15}^{+0.17}$ & \nodata & \nodata & 4.1$_{-1.3}^{+1.8}$ & 32.33/22 & 7.2e-02 \\
23 & 2014-Jan-31 & 16:52:37.461 & 4.98$_{-0.11}^{+0.11}$ & \nodata & \nodata & 95$_{-17}^{+21}$ & 43.80/21 & 3.0e-03 \\
24 & 2014-Feb-08 & 05:49:29.848 & 4.16$_{-0.12}^{+0.13}$ & \nodata & \nodata & 9.4$_{-2.2}^{+2.9}$ & 36.42/24 & 5.0e-02 \\
25 & 2014-Oct-18 & 02:49:17.710 & \nodata & \nodata & \nodata & \nodata &  \nodata & \nodata \\
26 & 2014-Oct-27 & 03:14:46.862 & 4.15$_{-0.21}^{+0.23}$ & \nodata & \nodata & 5.6$_{-2.2}^{+3.6}$ & 16.91/22 & 7.7e-01 \\
27 & 2015-May-07 & 12:41:40.415 & 3.61$_{-0.13}^{+0.14}$ & \nodata & \nodata & 1.8$_{-0.6}^{+0.8}$ & 29.83/27 & 3.2e-01 \\
\enddata

\tablenotetext{*}{The \kw\ trigger time after corrections for the light propagation to the Earth are applied.}
\tablenotetext{**}{Lower limits.}

\end{deluxetable*}

To recover spectrum of accelerated electrons from photon spectrum we used collisional thick-target model \citep{Brown1971} with power-law spectrum of nonthermal electrons (brmThickPL):

\begin{equation}
\label{eq_brmThickPL}
F(E) = \begin{cases} 0 & E < E_{cut,low} \\
\propto A E^{-\delta} & E_{cut,low} \le E \le E_{cut,high} \\
0 & E_{cut,high} < E \end{cases}
\end{equation}
where $A$ is the electron flux in electron~keV$^{\rm -1}$~s$^{\rm -1}$.
This model gives a simple relationship between the electron and photon power-law indices $\gamma$=$\delta-1$ \citep{Brown1971, SomovSyr1976}.

For a fraction of both CSFs and reference flares the spectral break is observed and brmThickPL model is not applicable in these cases. In principle, some spectral flattening may be caused by instrumental pile-up effect, thus we performed the Monte-Carlo modeling which revealed that for the \kw\ spectra this effect becomes significant for count rates $\geq$5$\times$10$^4$ counts/s (dead-time corrected), but among both CSFs and reference flares used for multichannel fitting there are no such intense events. A spectral flattening at low energies may also be induced by several physical effects: photospheric albedo \citep{Kontar2006}, non-uniform target ionization \citep{Holman2011} and energy losses associated with a return current \citep{Holman2011}. These effects give spectral breaks at lower energies of the HXR nonthermal spectra, while $E_{\rm break,ph}$$\geq$60~keV for all CSFs. Thus, we conclude that spectral flattening at low energies is caused by non-power-law spectrum of accelerated electrons and in addition to brmThickPL model we use collisional thick-target model with broken power-law spectum of nonthermal electrons (brmThick2PL):

\begin{equation}
\label{eq_brmThick2PL}
F(E) = \left\{
\begin{array}{ll}
0 & E < E_{cut,low}\\
\propto E^{-\delta_1} & E_{cut,low} \le E \le E_{break,el} \\
\propto E^{-\delta_2} & E_{br,e} \le E \le E_{cut,high} \\
0 & E > E_{cut,high},
\end{array} \right.
\end{equation}

To minimize potential contributions from photospheric albedo, non-uniform target ionization, and return current to the spectrum formation of reference flares we excluded 9 flares from the reference group where $E_{\rm break,ph}$$\leq$60~keV. To fairly compare the electron fluxes in all cases we decided to fix $E_{\rm cut,low}$ at 10~keV, which is approximately the lower bound for nonthermal electrons found in CSFs case studies \citep{Fl_etal_2011, Fl_etal_2016, Motorina}.

\begin{figure*}\centering
\includegraphics[width=1\textwidth]{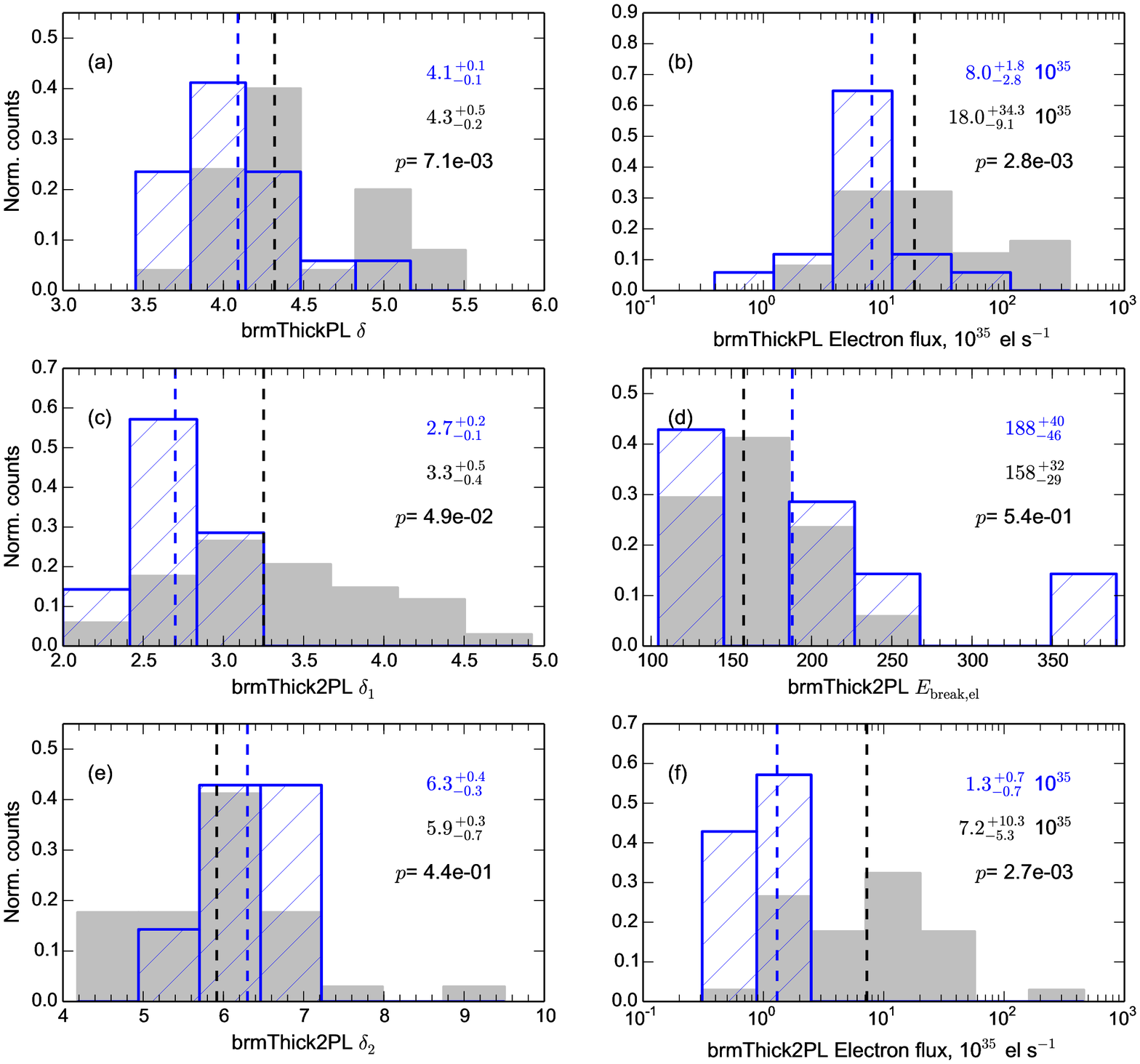}
\caption{\label{HXR_dist_el} Parameter distributions for electron spectra obtained from thick-target model. Bin heights are normalized to total number of events in each group. Blue hatched histograms refer to cold flares, grey histograms -- to reference group. Median values and 0.5 probability ranges are presented on each plot for cold flares and for reference group in blue and black letters respectively. Black "p"\ indicates two-sided p-value from Kolmogorov-Smirnov test in suggestion that distributions of a given parameter for cold flares and reference flares are equal. (a) brmThickPL electron spectral index, $\delta$, distributions; (b) brmThickPL electron flux distributions in 10~keV--40~MeV range; (c) brmThick2PL electron spectral index in lower energy range, $\delta_1$, distributions; (d) brmThick2PL break energy of electron spectrum, E$_{break,el}$, distributions; (e) brmThick2PL photon spectral index in higher energy range, $\delta_2$, distributions; (f) brmThick2PL electron flux distributions in 10~keV--40000~MeV range.
}
\end{figure*}

For some events fitting by brmThick2PL model gave unstable solutions and high correlations between parameters. This means, that the data agree with a broad range of fitting parameters, and we should choose the most physically valid ones among them. Thus, in those cases we froze high energy spectral index  according to thick-target model relationship between the photon and electron spectral indices $\delta_2=\gamma_2$+1 \citep{Brown1971, SomovSyr1976}. In some cases spectral steepening at energies $\geq$200~keV was very sharp and number of counts in this region was not enough to determine $\delta_2$, then we determined just lower limits of $\delta_2$.

For all events fitted with the photon PL model, the brmThickPL model was used, and most events with the energy break in the photon spectrum also have a break in the electron spectrum, except  1 CSF, 2002-Aug-20,  which was fitted by the 2PL model in the photon domain, while with a single brmThickPL model in the electron domain.

Distributions of the spectral parameter  $\delta$ and the electron flux for brmThickPL model are presented in Table~\ref{table_hxr_brm} and Figure~\ref{HXR_dist_el}(a--b). Likewise the photon spectral index, the electron spectral index is harder for CSFs, the  median value is 4.1, than for the reference flares, the  median value equals to 4.3, and this difference is rather significant: the probability that $\gamma$  distributions for CSFs and the reference group match is only 0.7~\%. The electron flux for CSFs is weaker than for the reference set, the median values are 8.0$\times$10$^{\rm 35}$~el~keV$^{\rm -1}$~$s^{\rm -1}$ and 18.0$\times$10$^{\rm 35}$~el~keV$^{\rm -1}$~$s^{\rm -1}$, respectively. The electron flux ranges from $\sim$0.4 to $\sim$300$\times$10$^{\rm 35}$~el~keV$^{\rm -1}$~$s^{\rm -1}$ for CSFs, and from $\sim$1 to  $\sim$300$\times$10$^{\rm 35}$~el~keV$^{\rm -1}$~$s^{\rm -1}$ for the reference group.

Distributions of the spectral parameters $\delta_1$, $E_{\rm break,el}$, $\delta_2$, and the electron flux for the brmThick2PL model are presented in Table~\ref{table_hxr_brm} and Figure~\ref{HXR_dist_el}(c--f). The median values of the electron low energy power-law index $\delta_1$ for CSFs and the reference group are 2.5 and 3.2, respectively, which is very close to the median values for $\gamma_1$ in Figure~\ref{HXR_dist_phot}. A probable reason why $\delta_1$ and $\gamma_1$ strongly deviate from the expected relationship $\gamma_1$=$\delta_1$-1 for both groups can be a large contributions from higher energy electrons to the  photon spectrum at lower energies. Distributions of the electron spectrum break energies $E_{\rm break,el}$ are presented in Figure~\ref{HXR_dist_el}(d). The median values for two groups are close to each other: 188~keV for CSFs, 158~keV for the reference group. Probability of distribution equality is large; p=54~\%. Distributions of the electron high-energy power-law index $\delta_2$ demonstrate that the reference flares might be slightly harder than CSFs, but this difference is insignificant, p=44~\%. The median value of CSF electron flux is much smaller than that for the reference group: 1.3$\times$10$^{35}$~el~keV$^{-1}$~s$^{-1}$ vs. 7.2$\times$10$^{35}$~el~$keV^{-1}$~$s^{-1}$. Probability that those distributions for CSFs and the reference group are equal is low, p=0.3~\%.

\subsection{\label{HXR_dur} Cold Solar Flare Timescales in HXR}

\begin{figure}[!h]\centering
\includegraphics[width=0.5\textwidth]{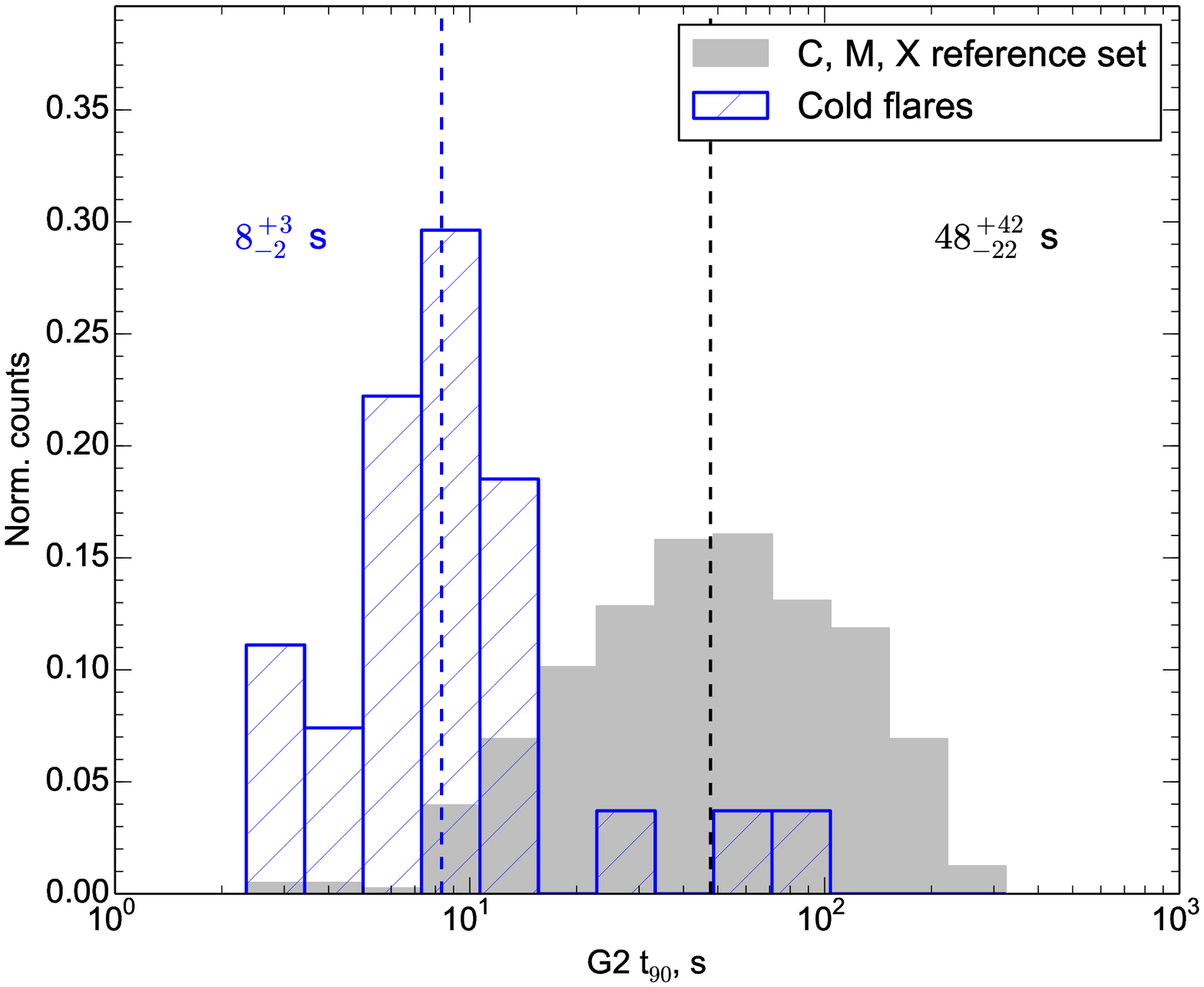}
\caption{\label{t90_dist} HXR duration estimation, $t_{90}$, distribution defined on G2 channel. Bin heights are normalized to total number of events in each group. Blue shaded bins refer to cold flares, grey bins refer to reference set of C-, M- and X-class flares. Dashed lines indicate median values for cold flares (blue) and reference set (black), labels refer to median values and 0.5 range for cold events (blue) and reference set (black).}
\end{figure}

The duration of the  impulsive flare phase in HXR was estimated using $t_{90}$ in G2 channel (see Section~\ref{HXR_vs_SXR} and Figure~\ref{DeltaGOES_ex}).
Values of $t_{90}$ for CSFs are listed in Table~\ref{table_sxr_hxr} and distributions of $t_{90}$ for the reference group and CSFs are presented in Figure~\ref{t90_dist}. It is apparent that CSFs are significantly shorter than flares from the reference group: the median duration of CSFs is 8~s, while for the reference set is 48~s. Most CSFs have $t_{90}$ between 5 and 20~s, 5 events have $t_{90}$ between $\sim$2 and $\sim$5~s, while there are 3 relatively long outliers: the longest event is 2000-May-18, 22:59~UT, ($t_{90}\simeq$86~s), and the other two are flare 2000-Mar-18 ($t_{90}\simeq$23~s) and flare 2014-Oct-18 (t$_{90}\simeq$63~s).

\subsection{\label{MW_fit} Spectral Properties of the  \Mw\ Bursts}

\begin{figure*}\centering
\includegraphics[width=1\textwidth]{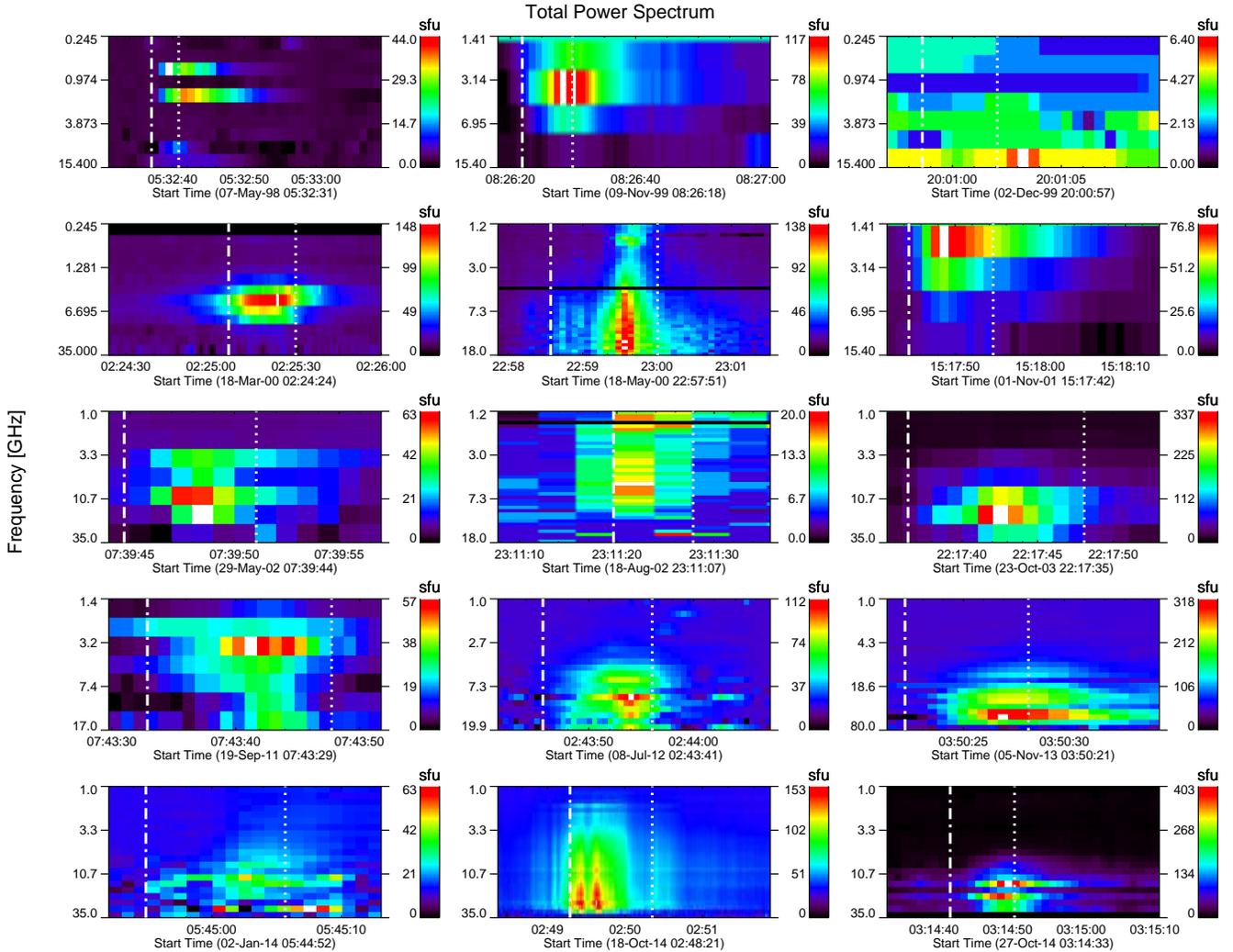}
\caption{\label{MW_dyn_fit}  \Mw\ dynamic spectra for 15 flares with successful spectral fits. Vertical dash-dotted line corresponds to the flare beginning in HXR range $t_{\rm 5}$, vertical dotted line corresponds to the flare ending in HXR range $t_{\rm 95}$ (see Section~\ref{HXR_vs_SXR}).
}
\end{figure*}

\begin{figure*}\centering
\includegraphics[width=1\textwidth]{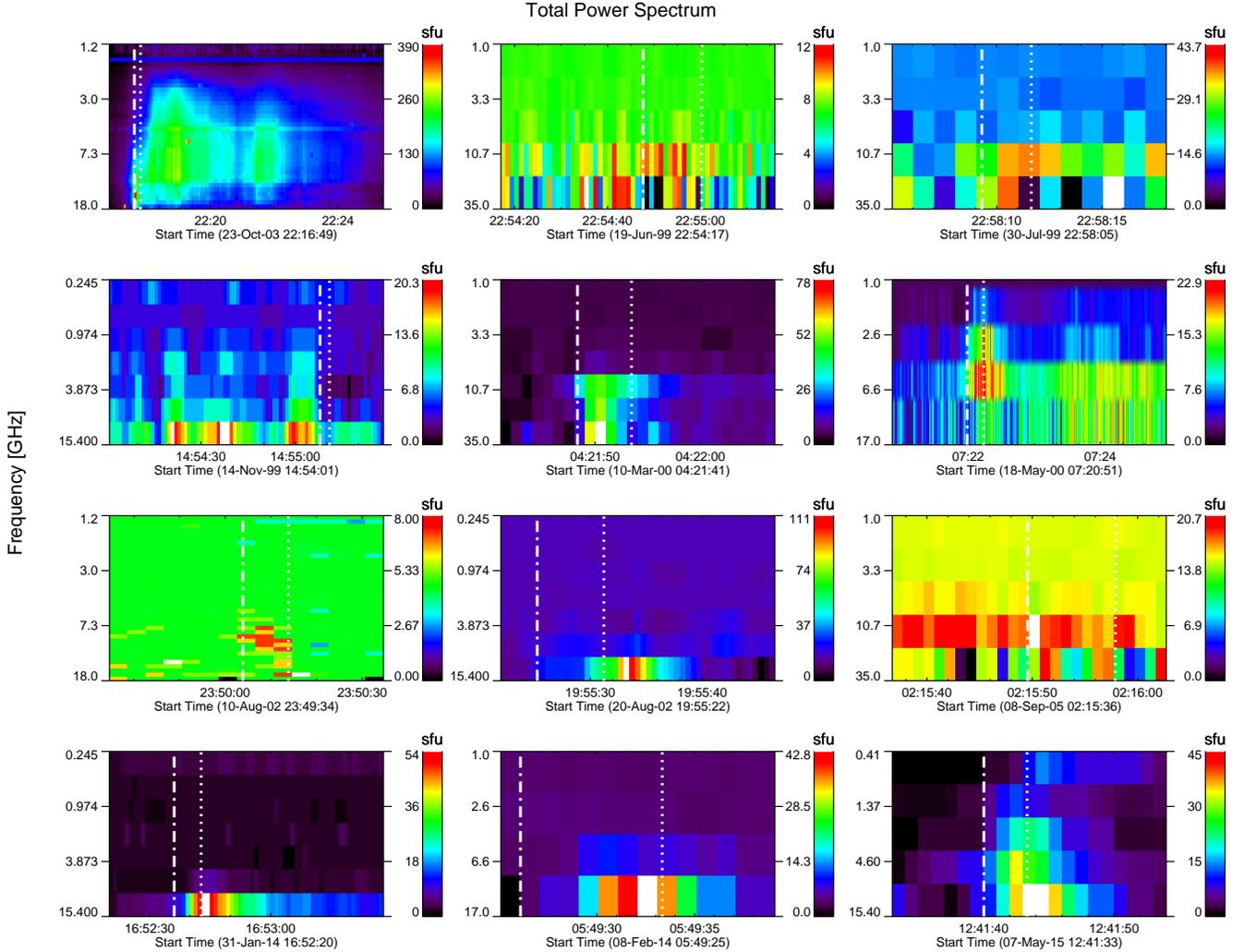}
\caption{\label{MW_dyn_nofit}  \Mw\ dynamic spectra for 11 flares for which spectral fitting didn't succeed and  long duration spectrum for 23-Oct-03, which was not used in \mw\ spectral parameter distributions. Vertical dash-dotted line corresponds to the flare beginning in HXR range $t_{\rm 5}$, vertical dotted line corresponds to the flare ending in HXR range $t_{\rm 95}$ (see Section~\ref{HXR_vs_SXR}).
}
\end{figure*}

\begin{deluxetable*}{lcccccccccc}
\tablecolumns{11}
\tablewidth{0pc}
\tablecaption{\label{table_mw}  \Mw\ properties of \csf s.
}
\tablehead{\colhead{N} & \colhead{Date}	& \colhead{$t_0$\tablenotemark{*}} & \colhead{MW instr.} & \colhead{$\Delta t$} & \colhead{$\langle S_{\rm peak}\rangle$} & \colhead{$\langle f_{\rm peak}\rangle$} & \colhead{$\langle\alpha_{\rm lf}\rangle$} & \colhead{$\langle\alpha_{\rm hf}\rangle$} & \colhead{$dt(f_{\rm low})$} & \colhead{$dt(f_{\rm high})$} \\
\colhead{} & \colhead{}	& \colhead{hh:mm:ss} & \colhead{} & \colhead{s} & \colhead{SFU} & \colhead{GHz} & \colhead{} & \colhead{} & \colhead{s} & \colhead{s}
}
\startdata
1 & 1998-May-07 & 05:32:37.072 & NoRP+RSTN & 16 & 30$\pm$10 & 1.3$\pm$0.8 & 1.9$\pm$0.5 & -1.3$\pm$0.6 & 2.0$\pm$1.0 & 1.5$\pm$1.0 \\
2 &	1999-Jun-19 & 22:54:49.788 & NoRP & \nodata & \nodata & \nodata & \nodata & \nodata & \nodata & \nodata \\
3 & 1999-Jul-30 & 22:58:09.675 & NoRP & \nodata & 40\tablenotemark{**} & 17\tablenotemark{**} & \nodata & \nodata & 1.5$\pm$1.0 & 1.5$\pm$1.0 \\
4 &	1999-Nov-09 & 08:26:21.703 & RSTN & 28 & 35$\pm$14 & 4.7$\pm$0.6 & 0.5$\pm$0.3 & -2.7$\pm$0.7 & \nodata & \nodata \\
5 & 1999-Nov-14 & 14:55:08.244 & RSTN & \nodata & \nodata & \nodata & \nodata & \nodata  & -4.5$\pm$1.0 & -5.6$\pm$1.0 \\
6 & 1999-Dec-02 & 20:01:00.012 & RSTN & 14 & 2.9$\pm$0.5 & 12$\pm$3 & 0.29$\pm$0.12 & -3$\pm$2v & 2.1$\pm$1.0 & 2.4$\pm$1.0 \\
7 & 2000-Mar-10 & 04:21:48.688 & NoRP+RSTN & \nodata & 110\tablenotemark{**} & 35\tablenotemark{**} & \nodata & \nodata & 1.3$\pm$1.0 & 1.5$\pm$1.0 \\
8 & 2000-Mar-18 & 02:25:10.567 & NoRP+RSTN & 53 & 166$\pm$72 & 6.3$\pm$1.7 & 3.1$\pm$0.9 & -3.4$\pm$1.6 & 0.0$\pm$1.0 & -1.8$\pm$1.0 \\
9 & 2000-May-18 & 07:21:59.706 & NoRP & \nodata & \nodata & \nodata & \nodata & \nodata & 4.9$\pm$1.0 & 110.8$\pm$1.0 \\
10 & 2000-May-18 & 22:59:39.777 & NoRP, OVSA & 67 &  67$\pm$24 & 13$\pm$0.8 & 1.19$\pm$0.29 & -5.1$\pm$1.7 & 17.4$\pm$1.0 & 12.7$\pm$1.0 \\
11 & 2001-Oct-12 & 07:40:31.941 & \nodata & \nodata  & \nodata & \nodata & \nodata & \nodata & \nodata & \nodata \\
12 & 2001-Nov-01 & 15:17:42.772 & RSTN & 11 & 52$\pm$20 & 1.43$\pm$0.05 & 2.6$\pm$0.8 & -1.7$\pm$0.6 & 5.0$\pm$1.0 & 2.7$\pm$1.0 \\
13 & 2002-May-29 & 07:39:46.864 & NoRP+RSTN & 7 & 36$\pm$19 & 14.4$\pm$12.0 & 1.4$\pm$1.0 & -2.5$\pm$1.7 & 1.1$\pm$1.0 & 1.0$\pm$1.0 \\
14 & 2002-Aug-10 & 23:50:09.293 & OVSA & \nodata & 5\tablenotemark{**} & 18\tablenotemark{**} & \nodata & \nodata & \nodata & \nodata \\
15 & 2002-Aug-18 & 23:11:19.740 & NoRP+RSTN & 34 & 15$\pm$7 & 9.4$\pm$2.5 & 1.3$\pm$0.3 & -2.6$\pm$1.5 & 3.6$\pm$1.0 & 0.1$\pm$1.0 \\
   &             &              &  OVSA &   &  &   &   &   &   &   \\
16 & 2002-Aug-20 & 19:55:28.476 & RSTN & \nodata & 110\tablenotemark{**} & 15.4\tablenotemark{**} & \nodata & \nodata & \nodata & \nodata \\
17 & 2003-Oct-23 & 22:17:39.620 & NoRP+RSTN  & 13 & 282$\pm$176 & 20$\pm$3 & 3.2$\pm$0.6 & -3.2$\pm$2.3 & 74.8$\pm$1.0 & 2.1$\pm$1.0 \\
   &             &              &  OVSA &   &  &   &   &   &   &   \\
18 & 2005-Sep-08 & 02:15:49.996 & NoRP & \nodata & \nodata & \nodata & \nodata & \nodata  & \nodata & \nodata \\
19 & 2011-Sep-19 & 07:43:40.791 & NoRP+RSTN  & 16 & 23$\pm$11 & 7$\pm$6 & 3.5$\pm$1.7 & -0.9$\pm$0.5  & 0.0$\pm$1.0 & 1.8$\pm$1.0 \\
20 & 2012-Jul-08 & 02:43:50.647 & NoRP+SRS & 8 & 60$\pm$22 & 12.1$\pm$1.8 & 2.4$\pm$0.7 & -1.3$\pm$0.7 & 1.6$\pm$1.0 & 0.2$\pm$1.0 \\
   &             &              &  +RSTN &   &  &   &   &   &   &   \\
21 & 2013-Nov-05 & 03:50:24.588 & NoRP+SRS & 11 &  202$\pm$71 & 23.1$\pm$2.3 & 1.96$\pm$0.27 & -3.0$\pm$0.7 & 5.4$\pm$1.0 & 1.3$\pm$1.0 \\
   &             &              &  +RSTN &   &  &   &   &   &   &   \\
22 & 2014-Jan-02 & 05:45:01.390 & NoRP+SRS & 9 & 22$\pm$7 & 12.7$\pm$0.6 & 1.8$\pm$0.8 & -2.6$\pm$1.4 & 5.8$\pm$1.0 & 3.1$\pm$1.0 \\
23 & 2014-Jan-31 & 16:52:37.461 & RSTN & \nodata & 54\tablenotemark{**} & 15.4\tablenotemark{**} & \nodata & \nodata & 7.7$\pm$1.0 & 6.5$\pm$1.0 \\
24 & 2014-Feb-08 & 05:49:29.848 & NoRP & \nodata & 55\tablenotemark{**} & 17\tablenotemark{**} & \nodata & \nodata & -0.4$\pm$1.0 & 0.1$\pm$1.0 \\
25 & 2014-Oct-18 & 02:49:17.710 & NoRP+BBMS & 195 & 69$\pm$37 & 8.1$\pm$0.4 & 3.3$\pm$0.5 & -1.3$\pm$0.3  & 0.8$\pm$1.0 & 0.5$\pm$1.0 \\
26 & 2014-Oct-27 & 03:14:46.862 & NoRP+SRS & 16 & 175$\pm$95 & 13.3$\pm$2.4 & 4.9$\pm$2.9 & -1.2$\pm$0.5 & 6.0$\pm$1.0 & 3.4$\pm$1.0 \\
27 & 2015-May-07 & 12:41:40.415 & RSTN+KMAS & \nodata & 45\tablenotemark{**} & 15.4\tablenotemark{**} & \nodata & \nodata & 1.6$\pm$1.0 & 3.2$\pm$1.0 \\
\enddata
\tablenotetext{*}{The \kw\ trigger time after corrections for the light propagation to the Earth are applied.}
\tablenotetext{**}{Lower limits.}

\end{deluxetable*}

\begin{figure*}\centering
\includegraphics[width=1\textwidth]{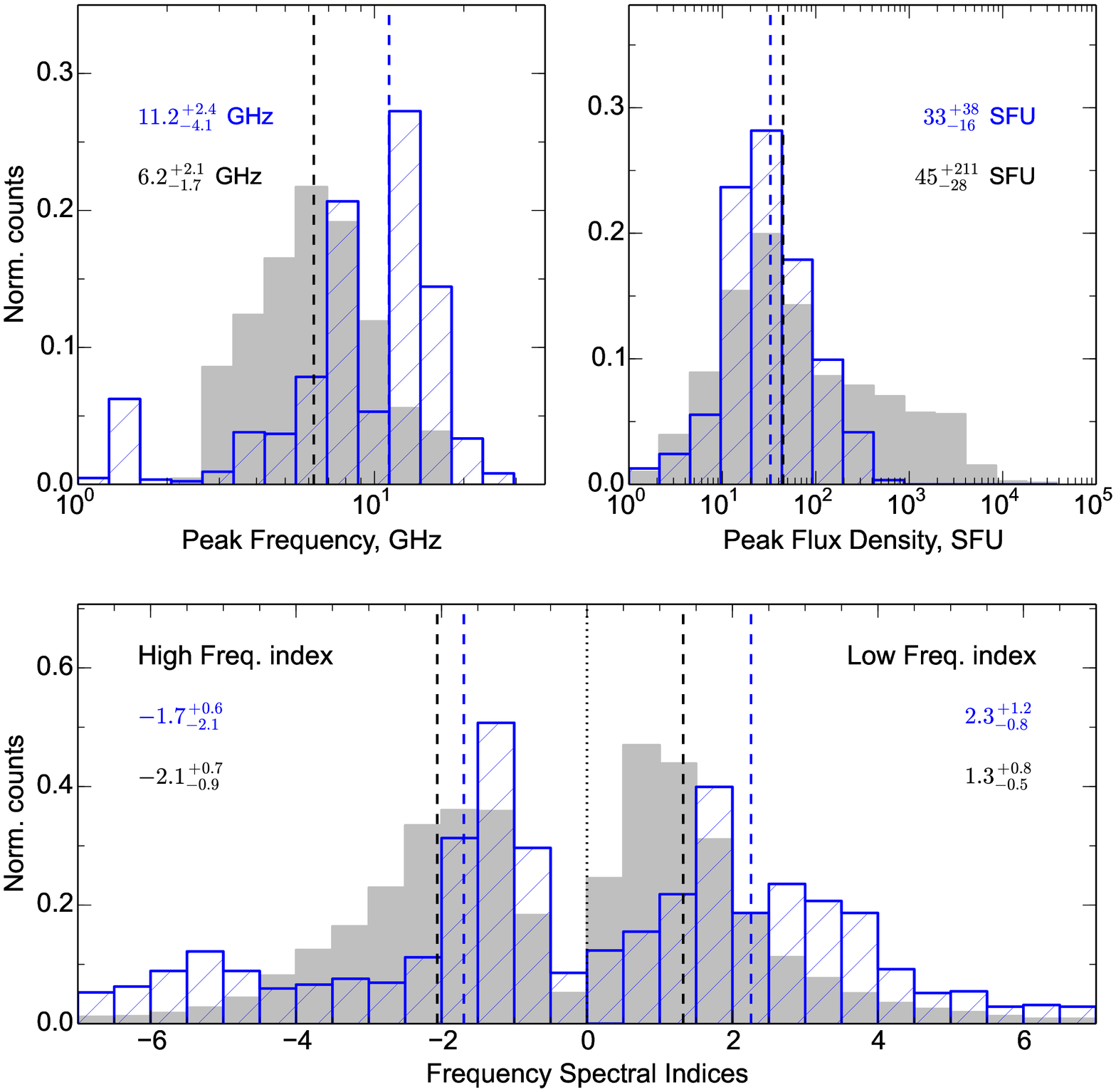}
\caption{\label{MW_specpars}  \Mw\ distributions of spectral parameters obtained on each time interval during each peak of \mw\ burst. Bin heights are normalized to total number of entries in each group. Top left: peak frequency distribution for \csf\ s (shaded blue) and reference group (grey); top right: peak flux density distribution for \csf\ s (shaded blue) and reference group (grey); bottom: spectral indices distributions for \csf\ s (shaded blue) and reference group (grey). Median values and 0.5 probability ranges are presented on each plot.
}
\end{figure*}

\begin{figure*}\centering
\includegraphics[width=1\textwidth]{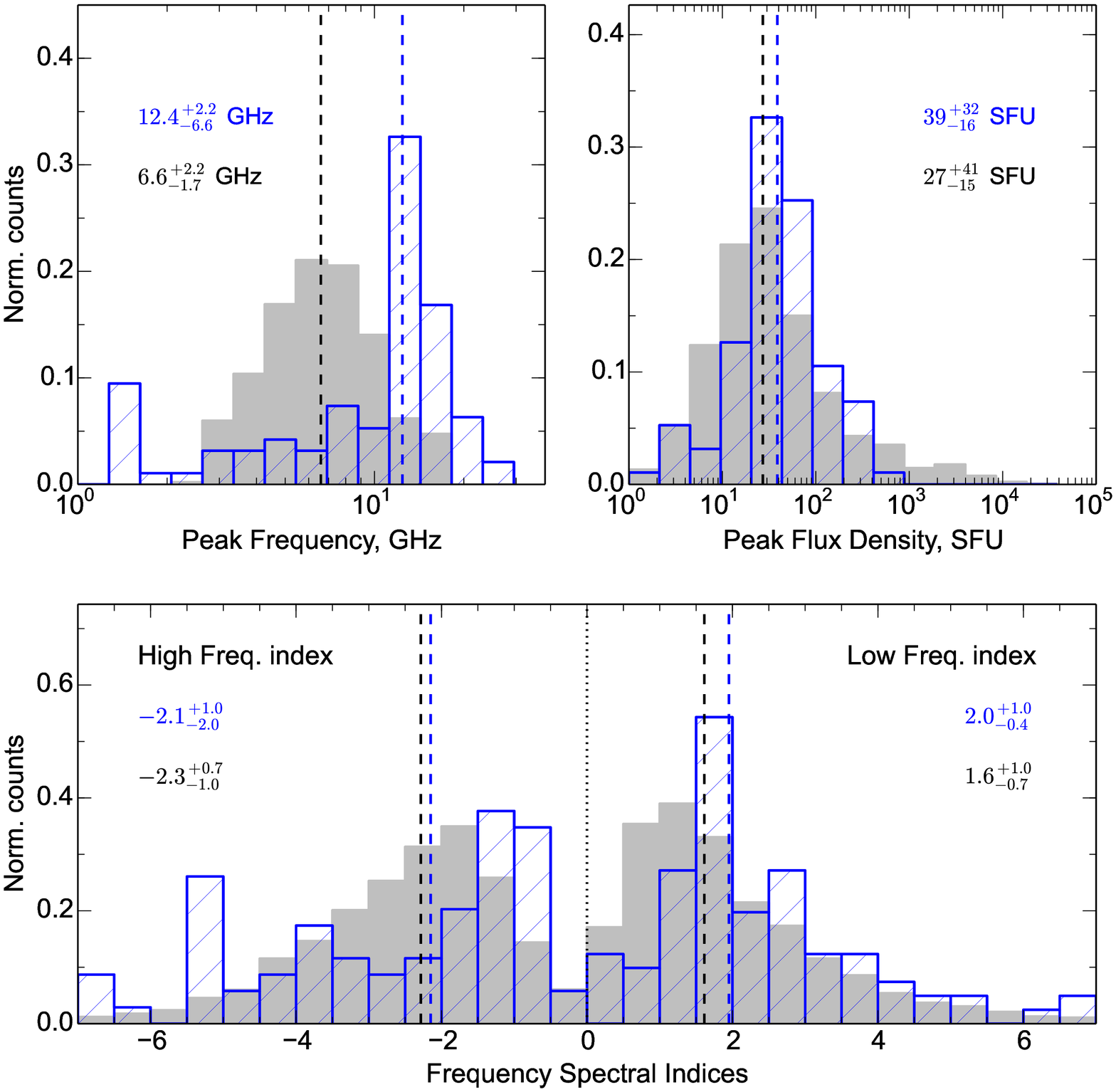}
\caption{\label{MW_specpars_5points_allpeak} \Mw\ distributions of spectral parameters obtained on 5 time intervals during each peak of \mw\ burst. Bin heights are normalized to total number of entries in each group. Top left: peak frequency distribution for \csf\ s (shaded blue) and reference group (grey); top right: peak flux density distribution for \csf\ s (shaded blue) and reference group (grey); bottom: spectral indices distributions for \csf\ s (shaded blue) and reference group (grey). Median values and 0.5 probability ranges are presented on each plot.
}
\end{figure*}

Given that the data files (IDL sav files) were created in a format identical to the OVSA data files, we took advantage of having the OVSA software from the SSW library. Specifically, the composite OVSA-like sav files have been read by \verb"OVSA_Explorer" widget, which has all functionality for data handling including the background subtraction and sequential spectral fitting.

Here we are only interested in the \mw\ bursts produced by gyrosynchrotron (GS) emission of energetic electrons, but ignore any component that could be attributed to a coherent plasma emission. GS spectrum $S(f)$ is characterized by a maximum flux density $S\p$ at a frequency $f\p$ and  two spectral slopes $\alpha_{\rm lf}$ in the low frequency range and $\alpha_{\rm hf}$ in the high frequency range \citep{Stahli1989}. We performed spectral fitting with the \verb"OVSA_Explorer" built-in generic spectral function:

\begin{equation}
\label{Eq_mw_fit}
S=e^{A} f^{\alpha}\left[1-e^{-e^{B} f^{-\beta}}\right],
\end{equation}
where $f$ is the frequency in GHz, while $A$, $B$, $\alpha$, and $\beta$ are the free fitting parameters, which yield the physical parameters of interest. For example, the low-frequency spectral index $\alpha_{\rm lf} \equiv \alpha$, while the high-frequency spectral index  is $\alpha_{\rm hf} = \alpha-\beta$. The peak frequency $f\p$ and the flux density at the peak frequency $S\p$ are calculated via parameters of the function $S$ \citep[see][for more detail]{Nita_etal_2004}.


Since \mw\ data for different events  have dissimilar time cadences, for spectral parameter analysis we first brought all the data to the same cadence selected to be 1~s. To this end the multifrequency time profiles of three OVSA events available with the cadence of 5 or 4 seconds were interpolated to 1 second resolution using IDL \verb"interpolate" routine; in the remaining cases the available 1~s data were used, which implies that in case of NoRP data we used the 1~s background data rather than the flare mode data. However, when the high-frequency light curves were deemed critical for the fitting, we also added the time-averaged NoRP 80~GHz light curve available in the flare mode only to the background 1~s record.

Having the combinations of the \mw\ data recorded by different instruments raises several problems with the data handling and analysis. Three main problems are (i) different flux calibrations at different instruments, (ii) different background levels, and (iii) possible clock errors and, thus, time scale mismatches. We dealt with all these issues individually for each event. As it was mentioned above, we used the NoRP clock as the reference time and adjusted clocks of other instruments using the lag-correlation between the light curves \citep[cf.,][]{Fl_etal_2016}. Total flux corrections were introduced based on comparison of the preflare signal levels, but this correction was not always successful; see prominent horizontal stripes in a few panels of Figure~\ref{MW_dyn_fit}, most notably, the upper left panel.

The frequency- and time- dependent background level was subtracted manually using the corresponding built-in functionality of the \verb"OVSA_Explorer", which allows defining a flat or polynomial background for each frequency channel.
In most cases it was sufficient to subtract a flat off-burst background level, although in some cases the background was time-dependent. In such cases the background was approximated by an appropriate polynomial.  In one case (2003-Oct-23), observed with both NoRP and OVSA, the short radio burst of interest occurred on top of a much more gradual broadband burst, whose dynamic spectrum is shown in the upper left panel of Figure~\ref{MW_dyn_nofit}. The (sub)burst of interest is seen at this dynamic spectrum as a short red dash at around 22:17:40~UT at the high-frequency end of the spectrum. One can notice a few more similar bright short subbursts later in the event; they are not artifacts as the same subbursts are detected by NoRP with higher time resolution; see the corresponding panel in Figure~\ref{MW_dyn_fit}. Thus, for our quantitative analysis of this event we use the 1~s cadence NoRP+RSTN data from which the gradual burst emission was subtracted\footnote{This bursts shares some apparent properties with the 2002-Mar-10 cold flare with delayed heating \citep{Fl_etal_2016} but studying the delayed component is beyond the scope of this statistical study.}. As a result, a data set with the background subtracted has been created and separately saved; the \mw\ dynamic spectra for all these events are given in Figures~\ref{MW_dyn_fit} and \ref{MW_dyn_nofit}.

This set of the background-subtracted dynamic spectra was used to perform sequential spectral fit for each 1~s time frame. The spectral fitting frequency range was chosen individually for each event to contain the main \mw\ component; we excluded low-frequency channels if a secondary, presumably coherent, component was present there, or high-frequency channels in case they were too noisy to aid the fitting. Although every attempt has been made to create accurate data files, the successful spectral fit was possible for only slightly more than half of all events (15 of 26); the corresponding dynamic spectra are gathered in Figure~\ref{MW_dyn_fit}. In the remaining 11 cases the fit failed entirely\footnote{The fit is possible for the left upper case, but this is the case treated as a background for a more impulsive subburst as explained above.}; see Figure~\ref{MW_dyn_nofit}. Visual inspection of  Figure~\ref{MW_dyn_nofit} suggests that fit might fail for the following reasons: (i) too small number of channels with a meaningful signal and (ii)  too high spectral peak frequency (15.4--35~GHz) such as a good fraction of the high-frequency \mw\ spectrum is outside the spectral coverage of the available instruments, which is the case for 5 of 11 events with no fit. The fit results for every successful time frame were saved in specifically designed OVSA med files, which are in fact a special form of the IDL sav files  \citep[see][for more detail]{Nita_etal_2004}\footnote{The created database of the composite \mw\ spectra is available at \url{http://www.ioffe.ru/LEA/SF\_AR/Radio.html}}.

It should be noted that even though the data were processed to create a database as uniform as possible, there are unavoidably some biases related to the individual instrument limitations and some anomalous features of the bursts. We will return to these biases later, when discussing the implications and significance of the results of our statistical analysis.

As a reference group in the  \mw\ domain we used the database from \cite{Nita_etal_2004} kindly provided by Dr. Gelu M. Nita. The \mw\ emission in our sample of events is undoubtedly the incoherent  \gyr\ emission even though the spectral peak frequencies vary in a wide range between $\sim1$~GHz and 35~GHz. The reference group of events, reported by \cite{Nita_etal_2004}, contains both \gyr\ and coherent bursts. \cite{Nita_etal_2004} found an empirical boundary at 2.6~GHz, which demarcates decimetric (D-type; often--coherent) and centimetric (C-type; mainly--\gyr)  bursts. In addition, a group of bursts with multiple spectral peaks in both centimeter and decimeter ranges was separated and called CD-type.
For a fair comparison between our event sample and the reference set, we only included the C-type bursts and the centimeter component of CD-type bursts in the statistical distribution of the reference group. This resulted in minor deviations in our numbers compared with those presented by \cite{Nita_etal_2004}, but those deviations are not statistically significant.

As \mw\ spectral parameters may vary significantly during the \mw\ burst duration \citep{Melnikov2008, Fl_etal_2016} for the  statistical study we used parameters obtained on each time frame with a successful fit during each \mw\ burst peak. Peak durations were estimated as time intervals during which the flux density at the local peak frequency is above 80~\% of the corresponding peak flux \citep[see][for more detail]{Nita_etal_2004}.

Burst averaged \mw\ spectral parameters for \csf s are presented in Table~\ref{table_mw}, while comparison of the spectral parameters of CSFs with the reference set for each time frame is illustrated in Figure~\ref{MW_specpars}. Even though we used essentially the same set of data, the displayed parameter distributions for the reference group differ from those described by \cite{Nita_etal_2004} because we display slightly different set of entries---those, which are required for our study.

The peak frequency distributions for the CSFs and the reference set are presented in Figure~\ref{MW_specpars}, top left. Histogram bins are the same as in \cite{Nita_etal_2004} for more straightforward comparison of the  results. For the  peak frequency distribution of CSFs, a small maximum near $\sim$1.5~GHz is observed followed by a broad minimum, then, after 2.6~GHz, the  distribution function begins to grow and has two more maxima, one near $\sim$6--9~GHz, which is close to the distribution maximum for the  reference group, and the second maximum between 11 and 18~GHz. Given that the histograms include the outcome of each time frame with a successful spectral fit, inputs from longer events have proportionally larger weights than those from shorter events. We checked and found that the two prominent peaks correspond to contributions from two long events: 2014-Oct-18 (the spectral peak frequency varies within 7--9~GHz) and 2000-May-18, 22:59~UT (the spectral peak frequency varies within 11--18~GHz); we will return to this bias later. The median spectral peak frequency for CSFs is 11.2~GHz, which is significantly larger than the median value for the reference group, 6.2~GHz. In fact, the true mismatch between  the spectral peak frequencies is even stronger. Indeed, we have already noticed from Figure~\ref{MW_dyn_nofit} that there are many (five, which is $\sim$20~\% of the total number of CSFs) events, whose spectral peak frequency is outside the available spectral range, i.e., at least above 15.4~GHz; such values, if properly added to the distribution, would further increase the median value of the spectral peak frequency of CSFs. However, those events do not contribute to the histograms because no fit was possible for them. Although we used NoRP data with wider frequency range extending up to 35~GHz in the background mode, the region 11--18~GHz lies within the OVSA coverage used by \cite{Nita_etal_2004}, thus we can conclude that the higher spectral peak frequencies in the CSF set are not due to a selection effect, but have a physical origin. 
It should, however, be noted that along with events having unusually high spectral peak frequencies, there are a few events with rather low peak frequencies, $f\p\sim2-3$~GHz, similar to the cold, tenuous flare described by \citet{Fl_etal_2011}.

\begin{figure}[!t]\centering
\includegraphics[width=0.5\textwidth]{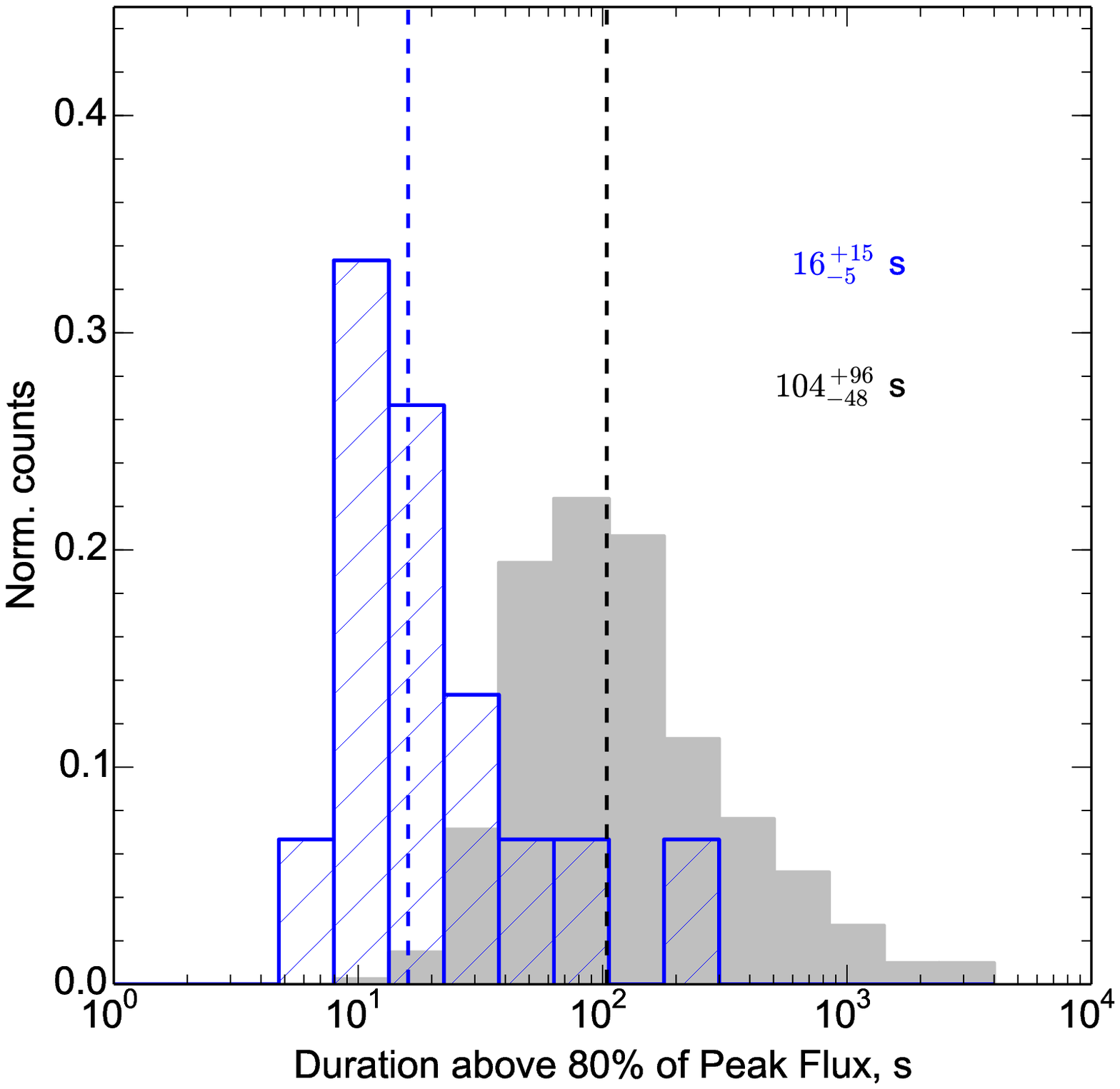}\\
\caption{\label{Duration_MW} \Mw\ duration distribution for \csf s (shaded blue) and reference flares (grey). Bin heights are normalized to total number of events in each group. Median values and 0.5 probability ranges are presented on each plot.}
\end{figure}

Distributions of the peak flux density are plotted in Figure~\ref{MW_specpars}, top right. These distributions have bell shapes with maxima between 20 and 50~sfu. The median values for the CSFs and the reference flares are close to each other, being equal to 33~sfu and 45~sfu, respectively. A minimum peak flux density for the time intervals where the spectral fitting for CSFs was successful is $\sim$4~sfu, while the maximum peak flux density is $\sim$600~sfu; the 50~\% probability range is from 17~sfu to 71~sfu. The reference group  50~\% probability range is between 17~sfu and 256~sfu;  no extremely intense burst with a peak flux density $>$1000~sfu was found among the CSFs.

Distributions of high- and low- frequency spectral indices for both groups are presented in Figure~\ref{MW_specpars}, bottom. For the high-frequency spectral index (negative region) the CSF median value is --1.7 and 50~\% probability range is within --3.8 and --1.1. The median value for the reference group is --2.1; thus, the CSFs are slightly harder, while 50~\% range for the reference flares is narrower and falls between --3.0 and --1.4. The low frequency spectral index (positive region) median value for the CSFs is 2.3, which is significantly larger than that for the reference group characterized by the median value of 1.3, although, like in the case of  high-frequency index distribution, the low-frequency slope distribution for CSFs has a wider 50~\% range, 1.5--3.5,  than the reference set distribution, 0.8--2.1.  Such wide ranges of the spectral indices for CSFs come from long distribution "tails"\ in the regions, where the spectral indices are large in absolute value. Some contribution to these tails can come from fitting artifacts, where the low- or high- frequency slope is only constrained by one spectral channel in cases when the spectral peak frequency is either very large or small. In such cases the corresponding spectral index is determined with a large error. In what follows we do not draw any physical conclusion based on the presence of these tails. If we neglect the second peak (tail), i. e. values $\leq$--4.5, of the high-frequency spectral index histogram, the median value for reference flares doesn't change sigficantly and becomes --2.0, while median for CSFs moves to --1.4 and coincides with the strongly pronounced maximum of CSF high frequency slope distribution.

The use of the entire flare duration in the statistics  described above is justified by the fact that there are cases with an extremely prominent spectral evolution, such as in a cold flare described by \citet{Fl_etal_2016}, where the spectral peak frequency varied within 1.5 orders of magnitude. In such cases, characterization of the event with only the peak value or a few `representative' time frames could be misleading. On the other hand, a `cold' flare reported  by \citet{Fl_etal_2011} did not show any spectral evolution, so a single time frame would be sufficient to fully describe its spectral shape. To balance such extremes in our statistical study, we complement the described treatment relying on all successfully fitted time frames, and so giving an enhanced weight to longer bursts, by a treatment that characterizes each distinct temporal peak by exactly five time frames independently on the flare duration; thus, giving equal weight to every peak. These five time frames are selected at the beginning and at the end of the peak, at the peak maximum and in the middles of the raising and declining phases.
The \mw\ spectral parameter distributions for the input specified this way  are presented in Figure~\ref{MW_specpars_5points_allpeak}.

Compared to Figure~\ref{MW_specpars}, the most significant change is observed in the
peak frequency distribution of CSFs; Figure~\ref{MW_specpars_5points_allpeak}, top left: the mid-frequency peak at $\sim$7~GHz has almost gone, while the high-frequency one has raised such as the
median value  for CSFs is now 12.4~GHz, which is significantly larger than the median value  for the reference group, 6.6~GHz.

Distributions of the peak flux density, Figure~\ref{MW_specpars_5points_allpeak}, top right, have changed only slightly. The median values for CSFs and reference flares are close to each other and are 39~SFU and 27~SFU, respectively.
Distributions of the high- and low- frequency spectral  indices for both groups are presented in Figure~\ref{MW_specpars_5points_allpeak}, bottom. For high frequency spectral index (negative region), the CSF median value is --2.1, and 50~\% probability range lies between --4.1 and --1.1. The median value for the  reference group is --2.3;  50~\% range for the reference flares is narrower and falls between --3.3 and --1.6. Though the  medians for the CSFs and reference group became closer to each other, the maximum between --2.0 and 0.0 for CSFs distribution still remains, thus the conclusion that there are significantly harder events among the CSFs is confirmed. The median value of the low-frequency spectral index (positive region) of CSFs is 2.0, which is larger than for the reference group characterized by the median value of 1.6 and coincides with the  maximum of the  CSFs distribution. Likewise the high-frequency spectral index distribution, the low-frequency one has a wider 50~\% range, 1.6--3.0, for the CSFs than for the reference set distribution, 0.9--2.6. We attempted a few other selections of the time frames for the statistical analysis, but did not find any trend differing from those reported above.

\subsection{\label{MW_dur} Cold Solar Flare Timescales in \Mw s}

To characterize the timescales of the \mw\ bursts we used the same approach as \cite{Nita_etal_2004},  who calculated the  duration at the absolute peak frequency \citep[see][for more detail]{Nita_etal_2004}. For events with multiple temporal peaks the main peak was taken. This approach allows  using the built-in \verb"OVSA_Explorer" functionality for the peak duration determination and then to compare directly our results to those obtained by \cite{Nita_etal_2004}. An unavoidable down side of this approach is that it can only be applied to 15 of 26 flares, for which the spectral fits were obtained.
The results are listed in Table~\ref{table_mw} (column $\Delta$t) and presented in Figure~\ref{Duration_MW}. \Mw\ peak durations for CSFs extend from 7~s to 195~s, but most events last less than 30~s. 2000-Mar-18, 2000-May-18, 22:59~UT, and 2014-Oct-18 flares, the sameas in the HXR range, have a relatively longer duration. An estimated median value of the burst duration is 16~s. This is much shorter compared to the reference flares, for which the median value\footnote{The value of 24~s in Table~2 in \citet{Nita_etal_2004} is incorrect likely due to a typo.} is 104~s and the durations extend up to thousands of seconds.

\subsection{\label{HXR_vs_MW} Relationships between X-ray and \Mw\ Flare Parameters}

HXRs are often associated with the injected population of flare-accelerated nonthermal electrons, while the \mw s with the trapped component \citep{Kosugi1988}. Thus, the study of relative timing and relationships between HXR and \mw\ spectral characteristics can shed light on properties of both these important ingredients of solar flares and therefore conditions of nonthermal electron propagation in flaring loops.

\subsubsection{Time Delays between \Mw\ and HXR Emission}

\begin{figure}\centering
\includegraphics[width=0.5\textwidth]{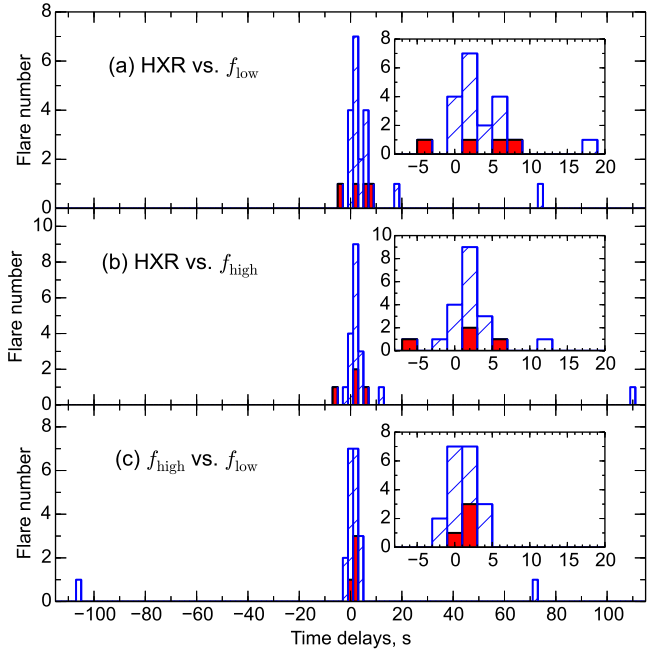}
\caption{\label{MW_KW_delay_dist} Time delays between HXR and \mw\ time profiles for \csf s. Red bins indicate the fraction of flares with only RSTN data in \mw\ range available. (a) Delays between time profile in G2 (HXR) and time profile at lowest \mw\ frequency, positive delay means that HXR emission is ahead. (b) Delays between time profile in G2 (HXR) and time profile at highest \mw\ frequency, positive delay means that HXR emission is ahead. (c) Delays between time profile at highest \mw\ frequency and lowest \mw\ frequency, positive delay means that \mw\ emission at highest frequency is ahead.}
\end{figure}

Delays of \mw\ emission relative to HXR emission often indicate trapping of nonthermal electrons in flaring loops. To calculate these delays we used the \kw\ HXR time profiles in G2 channel because this channel is not contaminated by thermal emission. In the \mw\ range, we took time profiles at the highest frequency, $f_{\rm high}$, where \mw\ burst was observed, which corresponds to optically thin \gyr\ emission. In addition to $f_{\rm high}$, we examined \mw\ emission delays relative to HXR emission at the lowest frequency with observable \gyr\ emission, $f_{\rm low}$. The frequencies $f_{\rm low}$ and $f_{\rm high}$ were selected for each flare via visual inspection.

The \kw\ time profiles were corrected for the light propagation time (see Section~\ref{HXR_vs_SXR}) and then interpolated to have the same time bins as the \mw\ time profiles. A lag-correlation between a HXR light curve and a \mw\ light curve  was calculated using IDL function \verb"c_correlate", then the correlation coefficient  dependence on the time delay was cubic-spline-interpolated with a step of 0.1~s. The time lag corresponding to the peak of this function was adopted as the time delay between the HXR and \mw\ light curves. Time delays between the HXR and \mw\ light curves were obtained for 21 of 26 events for which appropriate \mw\ data were available. For the remaining 5 events it was not possible to compute the delays because of the low signal-to-noise ratio and faults in \mw\ data.

The results are presented in Table~\ref{table_mw} and Figure~\ref{MW_KW_delay_dist}. Histogram bin width was selected to be 2~s, which is two times of NoRP background mode data resolution available for the majority of flares. For most CSFs \mw\ emission both on $f_{\rm low}$ and on $f_{\rm high}$ is delayed relative to HXR, for 4 flares \mw\ and HXR maxima coincide within 1~s, and for 2 flares light curves on $f_{\rm high}$ are ahead of HXR, but for one of these flares, only the RSTN data in \mw\ range are available, so the corresponding time delay may not be reliable. In most cases the delays of \mw\ emission do not exceed 10~s, the most frequent delays are between 1 and 3~s. One flare, 2000-May-18, 22:59~UT, based on NoRP has larger delays: 18.4~s on $f_{\rm low}$ and 12.8~s on $f_{\rm high}$.

\begin{figure}[!b]\centering
\includegraphics[width=0.5\textwidth]{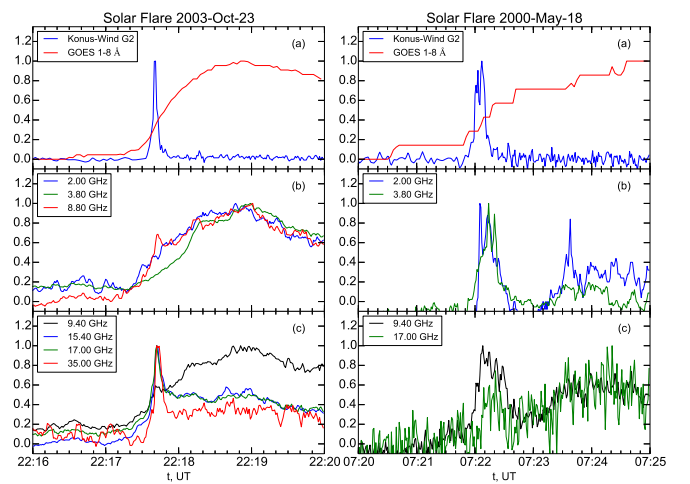}
\caption{\label{MW_KW_delay_ex} Example of solar flares with relatively large time delays between HXR and \mw\ emission.  Left panel:  solar flare 2003-Oct-23, t$_0$=22:17:42.391~s,~UT. Right panel: solar flare 2000-May-18, t$_0$=07:21:57.787~s,~UT. (a) \kw\ G2 channel and GOES 1--8~\AA\ light curves. (b) \Mw\ light curves on lower ($\leq$9~GHz) frequencies. (c) \Mw\ light curves on higher ($\geq$9~GHz) frequencies.}
\end{figure}

One flare, 2003-Oct-23, shows a smooth component that demonstrates a significant delay of the  \mw\ $f_{\rm low}$ emission relative to the  HXR emission, $\sim$75~s, Figure~\ref{MW_KW_delay_ex},~left. This is the flare, whose smooth component (top left dynamic spectrum in Figure~\ref{MW_dyn_nofit}) was subtracted as a background in our analysis performed so far. Now we consider the \mw\ emission as it is, subtracting the preflare background only. The time profiles in the \kw\ G2 channel and at high \mw\ frequencies ($\geq$9~GHz) demonstrate impulsive behavior, while the  \mw\ emission at the lower frequencies has a smooth time profile. This behavior is similar to that of the cold flare with a delayed heating described by \citet{Fl_etal_2016}. Given the high flux density and steep slopes of the spectrum at both low and high frequencies the emission is clearly nonthermal and, thus, this flare likely contains a nonthermal electron population trapped in a relatively large magnetic flux tube.

On the contrary, the 2000-May-18 07:21~UT, flare has a relatively large delay between the HXR and \mw\ time profiles at the high frequency, Figure~\ref{MW_KW_delay_ex},~right. For this event the light curves in the HXR G2 channel and \mw\ at lower frequencies (2.0 and 3.8~GHz) are highly correlated, but the emission on $f_{\rm high}$ is more gradual and its maximum is delayed relative to the impulsive phase by $\sim$110~s. This \mw\ burst is rather weak, $\sim10$~sfu, and, thus, the delayed component could be produced by the flare-heated plasma by either free-free or gyro- emission mechanism; cf. thermal flare reported by \citet{Fl_etal_2015}.

The \mw\ time profiles at $f_{\rm low}$ and $f_{\rm high}$ do not show any delay between each other for 7 events, while for most of the flares (10 cases) the light curves at the lower \mw\ frequencies are delayed relative to the higher-frequency ones. There are a few possible effects that can be responsible for such a delay. For example, it can be due to rise of the brightness temperature of the optically thick low-frequency emission due to nonthermal electron spectral hardening \citep{Melnikov_1994} or due to decrease of the free-free opacity provided by  plasma heating in the case of cold dense flares  \citep{Bastian_etal_2007}.

\subsubsection{Relationship between the flare duration in \Mw\ and HXR}

\begin{figure}[!h]\centering
\includegraphics[width=0.5\textwidth]{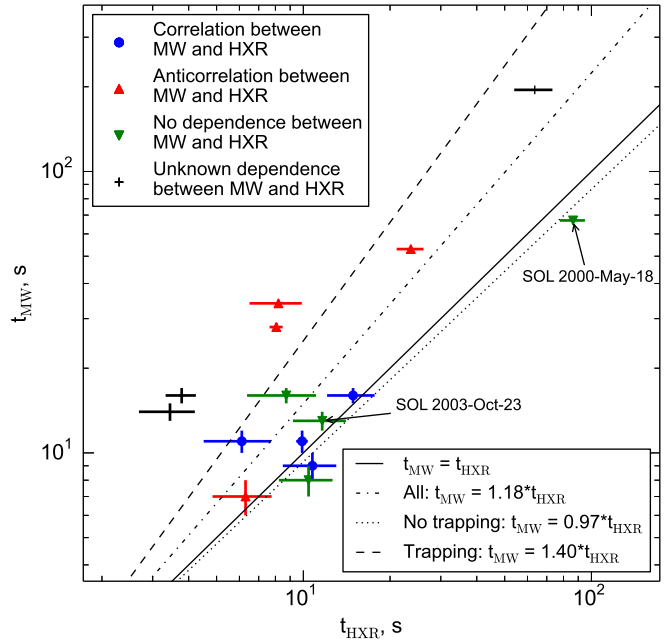}
\caption{\label{MW_KW_duration} Relationship between CSFs timescales in HXR and \mw. Solid line represents duration equality, dash-dotted line corresponds to linear regression between these timescales, dotted line reflects regression for the group of flares below the main regression (presumably without trapping), dashed line indicates regression for the flares above the main regression line (presumably with trapping). Explanation of ``colored'' event groups is given in Section~\ref{S_Spec_Ind_MX}. }
\end{figure}

Relationship between flare timescales in HXR and \mw\ ranges is presented in Figure~\ref{MW_KW_duration}. The events are split onto two groups: for one of them points are close to the solid line, which represents the equality of HXR and \mw\ durations, while the other group demonstrates the \mw\ burst significantly longer than the corresponding HXR burst. A simple interpretation of these trends is that the first group is composed of trapping-free events, while the trapping plays some role in the events from the second group.

\subsubsection{\Mw\ vs. HXR spectral indices}
\label{S_Spec_Ind_MX}

\begin{figure*}\centering
\includegraphics[width=1\textwidth]{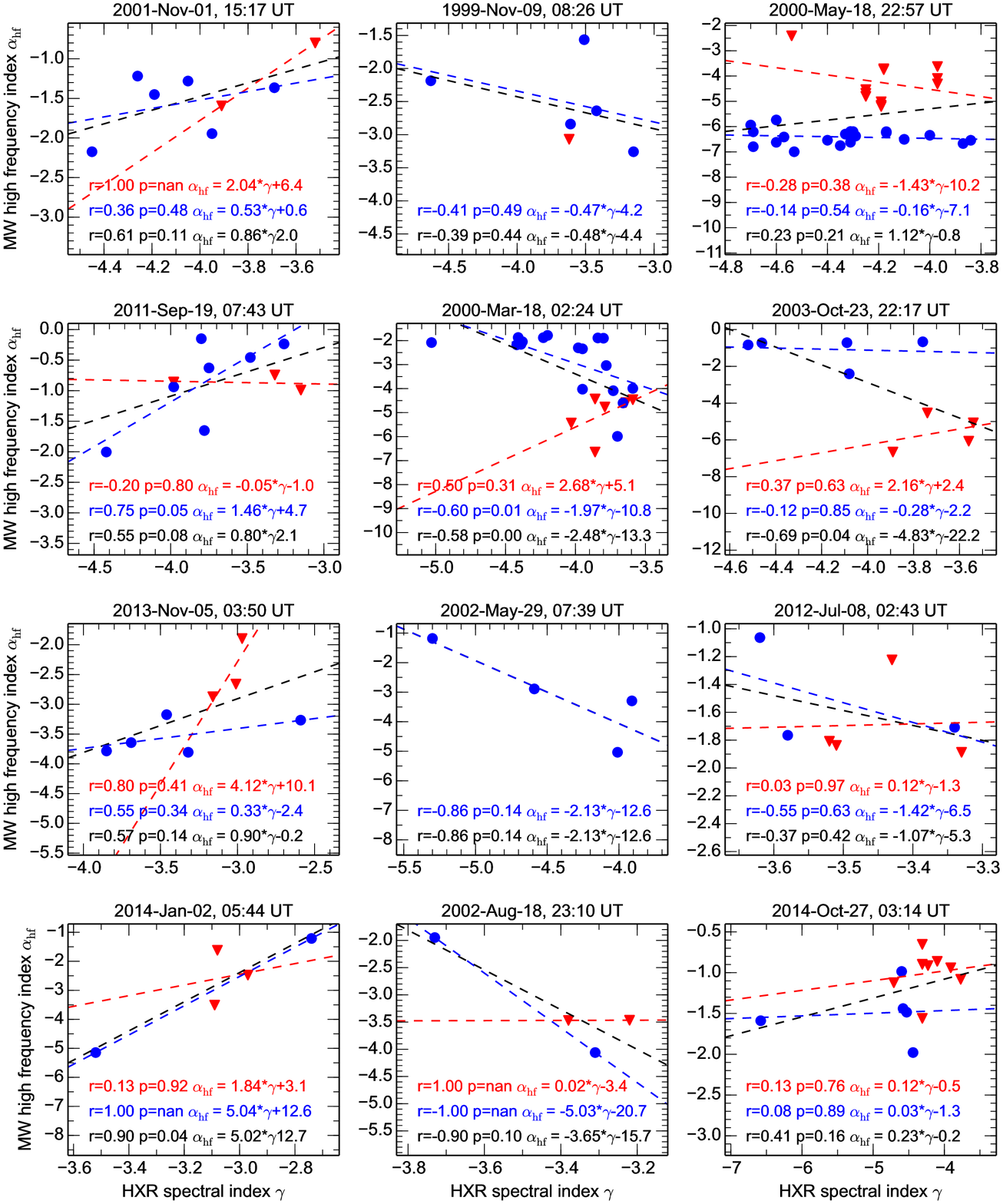}
\caption{\label{MWpow_vs_kwpow} Relationship between HXR spectral index $\gamma$ and \mw\ high frequency index $\alpha_{hf}$ obtained on each time interval during peak with successful \mw\ and HXR spectral fitting. HXR spectral indices with high temporal resolution were obtained using 3-channel fitting of \kw\ data (see Sec.~\ref{HXR_vs_SXR}). Red triangles and blue circles correspond to rise and decay phases of \mw\ emission correspondingly. Red labels indicate relationships between $\gamma$ and $\alpha_{hf}$ on the rise phase, blue labels indicate relationships on the decay phase, black labels indicate relationships during the whole flare.}
\end{figure*}

Here we compare the high-frequency \mw\ spectral indices $\alpha_{\rm hf}$, which correspond to the optically thin \gyr\ emission and, thus, closely linked with the  spectral indices of nonthermal electron distribution in the flaring loop, and the HXR spectral indices $\gamma$, which are associated with the spectral indices of injected electrons.

\Mw\ spectral parameters were obtained on 1~s time scales, but the multichannel HXR spectra with such time resolution were not available for the majority of flares, thus, the power-law indices in the HXR range, $\gamma$, were calculated using 3-channel fitting (see Section~\ref{HXR_vs_SXR}) on the time intervals in \kw\ G1, G2, G3 channels corresponding to the intervals with successful \mw\ fits after correction for the  propagation time. HXR fit results with fit probabilities $<$1~\verb"%" were discarded. Surprisingly, the scatter plot (not shown) of $\alpha_{\rm hf}$ vs $\gamma$ does not reveal any significant correlation between these two parameters; so we investigate this relationship on the event-by-event basis.

It is known \citep{Trottet1984, Meln_Magun_1998} that for some flares the HXR and \mw\ spectral indices behave consistently at the rise phase but show opposite trends at the decay phase, which has been interpreted as an outcome of Coulomb collisions of the trapped population of the nonthermal electrons with the ambient plasma particles. For this reason we consider the relationships between \mw\ high frequency spectral index, $\alpha_{\rm hf}$, and HXR index, $\gamma$, separately for the rise and decay phases of \mw\ emission.

The data suitable for comparison between $\alpha_{\rm hf}$ and $\gamma$ were obtained for twelve out of fifteen CSFs with successful \mw\ fits. Three flares were excluded because no triggered mode data were available in the G2 and G3 channels for the 2014-Oct-18 flare, while the high frequency spectral indices could not be obtained reliably for the  1998-May-07 and 1999-Dec-02 flares due to small number of the spectral data points or due to a  weak signal to the right of the turnover frequency. Time frames with weak signal or only one data point above the spectral peak frequency were excluded (recall, these same time frames contribute to the ``tails'' of $\alpha_{\rm hf}$ distributions as has been discussed in Section~\ref{MW_fit}).

The results of this analysis are presented in Figure~\ref{MWpow_vs_kwpow}. All CSFs in general follow the soft-hard-soft spectral evolution in HXR range, while according to their spectral evolution in \mw\ range and the relationship between $\gamma$ and $\alpha_{\rm hf}$, CSFs can be roughly categorized onto three groups. The first group (the first column in Figure~\ref{MWpow_vs_kwpow}) includes flares  2001-Nov-01, 2011-Sep-19, 2013-Nov-05, and 2014-Jan-02, for which the  correlation between $\alpha_{\rm hf}$ and $\gamma$ is observed. Of these four flares, the 2001-Nov-01 flare and 2013-Nov-05 flare show a slight correlation between the spectral indices in the \mw\ and HXR ranges  during both the rise and the decay phases, while for other two flares from this group no relationship during the rise phase is revealed. The second group (the second column in Figure~\ref{MWpow_vs_kwpow}) includes 1999-Nov-09, 2000-Mar-18, 2002-May-29, 2002-Aug-18 flares, which are characterized by an anticorrelation between $\alpha_{\rm hf}$ and $\gamma$ during the decay phase. One flare of this group, 2000-Mar-18, shows correlation between \mw\ and HXR indices during the rise phase, while for other three flares the corresponding regressions at the rise phase either do not show any clear trend or do not contain enough data points to draw a conclusion. The third group (the third column in Figure~\ref{MWpow_vs_kwpow}) includes 2000-May-18, 22:57~UT, 2003-Oct-23, 2012-Jul-08, and 2014-Oct-27 flares, which do not show any clear dependence between the \mw\ and HXR spectral indices during both the rise and the decay phases. An interesting feature of 2000-May-18, 22:57~UT and 2003-Oct-23 flares of this group is a striking difference between $\alpha_{\rm hf}$ during the rise and decay \mw\ phases. For the 2000-May-18, 22:57~UT flare the spectral indices are harder at the rise phase as compared to the decay phase, while for the 2003-Oct-23 flare, the \mw\ spectral indices are harder during the decay phase. Thus, we can conclude that for the flares from the second group and for one flare from the third group, the \mw\ spectrum hardening is observed. On the contrary, for flares from the first group and for one flare from the third group the \mw\ spectrum becomes softer during the flare.

There is a correspondence between the flare duration patterns revealed by Figure~\ref{MW_KW_duration} and the spectral evolution patterns identified in Figure~\ref{MWpow_vs_kwpow}, in terms of presence or absence of the trapping in the given event. Indeed, the events elongated along the $y=x$ duration equality line in Figure~\ref{MW_KW_duration} are mainly from columns 1 (three events) and 3 (three events) in Figure~\ref{MWpow_vs_kwpow} with only one event from column 2. Three other events from column 2 (showing a \mw\ spectrum hardening at the decay phase indicative of spectral evolution of the trapped component of nonthermal electrons) are those with noticeably longer duration in \mw\ than in the HXR domain, which confirms that trapping plays a role in these flares.

\subsubsection{Relationships between X-ray characteristics and \Mw\ peak frequencies}

\begin{figure}\centering
\includegraphics[width=0.5\textwidth]{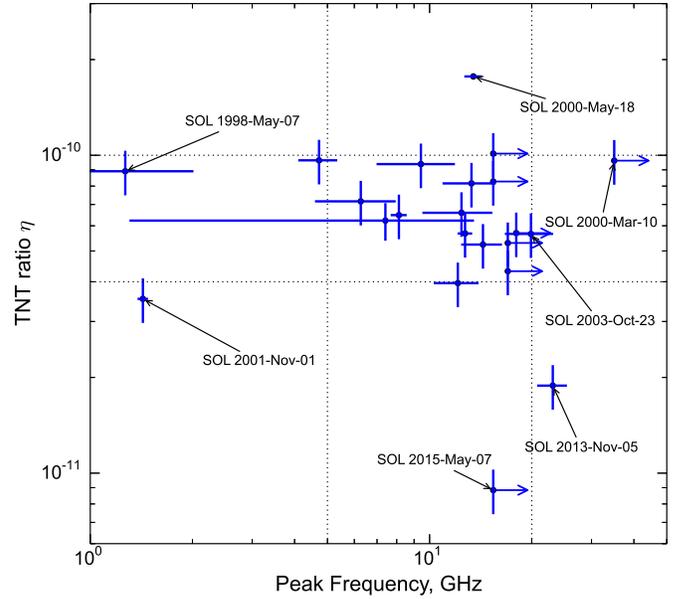}
\caption{\label{MWPF_vs_DeltaGOES} Relationship between $\Delta$GOES to \kw\ HXR peak count rate ratio and \mw\ peak frequency. \Mw\ peak frequencies were averaged over time frames with successful \mw\ fits for each flare and error bars refer to their standard deviations. For flares with fail fits where it was possible lower limits of peak frequency were estimated.}
\end{figure}

\begin{figure}\centering
\includegraphics[width=0.5\textwidth]{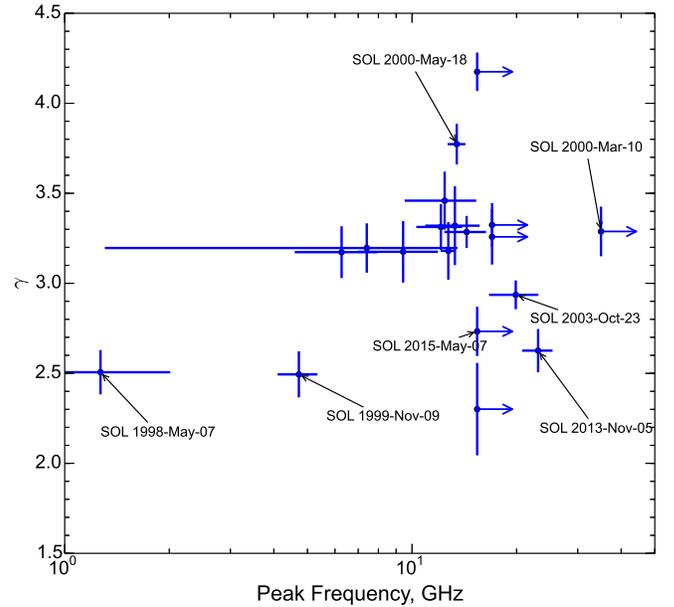}
\caption{\label{MWPF_vs_gamma} Relationship between photon power index in lower energy range $\gamma$ and \mw\ peak frequency. \Mw\ peak frequencies were averaged over time frames with successful \mw\ fits for each flare and error bars refer to their standard deviations. For flares with fail fits where it was possible lower limits of peak frequency were estimated.}
\end{figure}

In Section~\ref{MW_fit} we found that the CSFs have strikingly higher  spectral peak frequencies $f\p$ compared to the reference set of the \mw\ bursts. From this perspective it looks interesting to consider if any other CSF parameter correlates with the \mw\ spectral peak frequency.

To compare the \mw\ spectral peak frequencies obtained on individual time frames with X-ray parameters characterising the entire duration of the solar flare, the peak frequencies for flares with successful fits were averaged over time frames for each flare peak; the corresponding uncertainty of the spectral peak frequency was estimated as the standard deviation from this mean. For the `no-fit' flares, in some cases it was possible to specify a lower limit of the spectral peak frequency directly from the dynamic spectra given in Figure~\ref{MW_dyn_nofit}.

The relationship between the thermal-nonthermal (TNT) ratio $\eta=\Delta$GOES/(HXR peak count rate) and the mean \mw\ peak frequencies is displayed in Figure~\ref{MWPF_vs_DeltaGOES}. This plot shows, that most of the CSFs group between $\sim$5~GHz and $\sim$20~GHz in the peak frequency and between $\sim$4$\times$10$^{-11}$ and $\sim$1$\times$10$^{-10}$ in the TNT ratio (see dotted lines in the plot). Some flares fall outside these ranges. These are the 2000-May-18 flare, which has a rather high TNT  ratio, the  2001-Nov-01 flare with a low $\Delta$GOES to HXR peak count rate TNT ratio and a low mean peak frequency (1.4~GHz); the  2013-Nov-05 flare that demonstrates a low TNT ratio and an exceptionally high mean peak frequency, 22~GHz; and the  2015-May-07 flare with an exceptionally low TNT ratio.

The relationship between the photon power index in the lower energy range, $\gamma$ for PL model or $\gamma_1$ for 2PL model and the mean \mw\ peak frequencies, is presented in Figure~\ref{MWPF_vs_gamma}. The lower-frequency CSFs tend to be harder than the higher-frequency ones. There is no case, where a soft spectrum and a low spectral peak frequency would be present simultaneously.

\begin{figure}\centering
\includegraphics[width=0.5\textwidth]{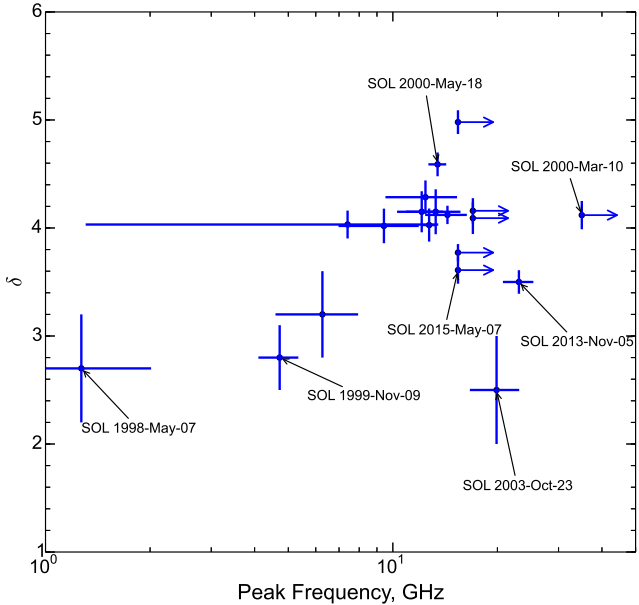}
\caption{\label{MWPF_vs_delta} Relationship between electron power index in lower energy range $\delta$ and \mw\ peak frequency. \Mw\ peak frequencies were averaged over time frames with successful \mw\ fits for each flare and error bars refer to their standard deviations. For flares with fail fits where it was possible lower limits of peak frequency were estimated.}
\end{figure}

The relationship between the electron power index in the lower energy range, $\delta$ for brmThickPL model or $\delta_1$ for brmThick2PL model and the mean \mw\ peak frequencies, is presented in Figure~\ref{MWPF_vs_delta}. There is a slight tendency that events with lower peak frequencies has harder $\delta$. Here we see the same trend as in the previous scatter plot of $\gamma$ vs $f\p$.

\section{\label{Discussion} Discussion}

In this study we have identified a statistically significant set of ``cold'' flares with a disproportionally weak thermal relative to the nonthermal emission. Although a few such cold flares were reported in a number of case studies, all those cases were selected subjectively without any formal criterion. Thus, to perform this statistical study, we have started with formulating such a formal criterion, which appears to be rather strict. Specifically, we identified a group of \eif s by selecting those events, in which the HXR burst started before or without the corresponding SXR GOES flare. Then, we computed a SXR enhancement during the HXR bursts and compared this enhancement for a referent set of bursts and \eif s. Finally, we selected the \eif s, which are also the outliers (far below `average' thermal response) in the scatter plots in Figure~\ref{DeltaGOES_instr}, to form our CSF list. Note that Figure~\ref{DeltaGOES_instr} contains many more outliers below the confidence interval, which are not \eif s according our selection criterion. We, however, did not add them to the list because for the non-\eif s, such a relatively weak GOES enhancement during the HXR burst could be an outcome of a strong pre-heating (i.e., the overall GOES enhancement is strong, but the enhancement due to nonthermal electrons is weak)---such flares, certainly, would not qualify as the cold ones. This selection yielded 27 CSFs, which we analyzed using X-Ray and \mw\ data and the cross-correlations between them.

The performed statistical study reveals significant differences between the CSFs and other flares. In the HXR domain the CSFs are harder, shorter, and weaker than the reference flares. In the \mw\ domain the CSFs are, however, not weaker than the referent bursts. Further, in the \mw\ domain the CSFs are shorter and harder at high frequencies than the reference ones, while steeper at the low frequencies. In addition, the CSFs often have a strikingly higher spectral peak frequency than the reference ones in the \mw\ domain. Nevertheless, we found that CSFs do not represent a uniform group of evens, but rather could be separated onto a few subclasses. In particular, some CSFs show signatures of the nonthermal electron trapping in a coronal flaring loop, while others do not show any trapping \citep[as in the main flaring loop in the flare reported by][]{Fl_etal_2016}; some flares demonstrate a break in the nonthermal electron energy spectrum, while others are consistent with a single power-law; some CSFs are likely produced in a dense source as in the cases reported by \citet{Bastian_etal_2007, Masuda2013}, while others -- in a tenuous one similar to the tenuous flare reported by \citet{Fl_etal_2011}.

Perhaps, the key to interpret these distinctions is the combination of a weaker intensity in HXR and normal intensity in \mw. Indeed, weaker HXR emission implies a weaker component of the nonthermal electrons accelerated in the flare. But, to produce a normal level of \mw\ \gyr\ emission by a weaker population of nonthermal electrons, the magnetic field must be accordingly higher than in the referent flare. The stronger magnetic field further implies a higher spectral peak frequency of \gyr\ emission as observed. Then, given that the magnetic field decreases with height in the corona, a strong magnetic field implies a reasonably low height of the radio source, thus, shorter flaring loops, and thus, shorter burst duration. These more compact loops represent  more uniform sources, i.e., they likely contain a narrower range of magnetic field strength than a bigger loop, which explains, why the low-frequency \mw\ slope is steeper in the CSFs compared with the referent ones. This simple interpretation cannot, however, clarify, what is the reason of harder spectra of nonthermal electrons in the CSFs compared with the referent ones. One possibility is that the  acceleration mechanism results in harder spectra in case of stronger magnetic field / more compact loops. An alternative is the spectral hardness does not systematically depend on these flare parameters, but in the flares that generate electrons with harder spectra, the thermal response is additionally reduced because now a bigger fraction of the nonthermal energy belongs to high-energy electrons, which deposit their energy deeper in the chromosphere, thus reducing the chromospheric evaporation and the thermal response.

However, even the relatively compact set of 27 CSFs demonstrates a considerable diversity of the properties, so that the CSFs can hardly be fully characterized by a ``mean'' CSF, but rather have to be further split onto different subclasses. This is particularly true for those CSFs showing a normal or low \mw\ spectral peak frequency: in such cases the magnetic field is also supposed to be normal or weak, so the picture drawn above should be patched or replaced, although the short duration still implies a reasonable compact flaring sources. We note, that in such cases, the nonthermal electron spectra are particularly hard, which might further confirm the role of the spectral hardness in the chromospheric evaporation / thermal response.
We note that having a low spectral peak frequency requires that both magnetic field and the plasma density are low. Indeed, a weak magnetic field combined with a high plasma density will result in a high spectral peak frequency due to the Razin-effect. Thus, the flares with a low spectral peak frequency are likely tenuous flares similar to that reported by \citet{Fl_etal_2011}.

Another property, which display different patterns within CSFs is the shape of the nonthermal electron energy spectrum. At this point we cannot draw any conclusion about significance of this finding given that in some cases the spectral break can be present in the spectrum, but not recovered in the fit due to insufficient statistics at the high-energy channels.

Finally, we found that the CSFs can be further divided onto two groups, depending on whether the nonthermal electron trapping in a coronal flaring loop (or loops) plays a role or not. This is vividly demonstrated by Figure~\ref{MW_KW_duration} showing the scatter plot of the \mw\ vs HXR burst duration. Roughly half of the events show equality of the durations (no trapping), while in the other half the \mw\ duration is a factor of two longer than the HXR duration (a noticeable trapping). This division is also confirmed by relationships of the spectral hardness in the \mw\ and HXR domains, Section~\ref{S_Spec_Ind_MX}.

\section{\label{Conclusions} Conclusions}

From the performed statistical analysis we conclude that the our identified set of 27 ``cold'' solar flares demonstrates properties statistically different from those of a referent (``mean'') flare. We found that the cold flares are typically shorter and harder that the referent flares; their HXR emission is weaker, while \mw\ emission is comparable to that of the mean flare. They are further different in the \mw\ domain as they have a steeper low-frequency spectral slope and (often) strikingly higher spectral peak frequency. From these findings we conclude that the cold flares are typically produced in more compact structures (presumably, short flaring loops) having stronger magnetic fields, than a mean flare. In addition, we found that the group of the cold flares is nonuniform, but rather can be further sub-categorized according to various properties (low or high spectral peak frequency, presence or absence of trapping effect etc). Given that all these flares demonstrate the weakest thermal response compared with a referent flare, we conclude that the described here ``cold'' flares offer, via the corresponding case studies, the cleanest way to study the electron acceleration in flares and the thermal plasma response driven by the nonthermal electron population.

We note that the presented here CSFs form a promising event list for future case studies given that the processes of particle acceleration and the thermal plasma response can be quantified much cleaner with the nonthermal-energy-dominated CSFs than with a `usual' flare, where the nonthermal and thermal energies are initially mixed up with unknown proportions. Such case studies will employ all available imaging and contextual data and also incorporate modeling to verify and refine interpretations formulated here based on the statistical analysis.

\acknowledgements

We are highly grateful to Dr. Valentin Pal'shin for initiation of this work and his help with \kw\ data. We thank anonymous referees who helped us to improve the paper.
We thank Dr. Dmitry Svinkin, Dr. Dmitry Frederiks and Mr. Mikhail Ulanov for useful advice concerning analysis of the \kw\ data, Prof. Astrid Veronig for productive discussion and Dr. Gelu Nita for providing \mw\ data from their paper and consultations. AL gratefully acknowledge support from RSF grant 17-12-01378. GF thanks  NSF grants  AGS-1250374 and AGS-1262772, NASA grant NNX14AC87G to New Jersey Institute of Technology and  RFBR grant 15-02-03835. NM acknowledge RFBR grants 15-02-03717 and 15-02-08028. AT and DZ acknowledge support from the Ministry of Education and Science of Russian Federation and from Siberian Branch of the Russian Academy of Sciences (Project II.16.3.2) and from the Program of basic research of the RAS Presidium No.28. Experimental data were obtained using the Common Use Centre "Angara" equipment.





\bibliography{ColdSF_stat,cold_flare_ref,fleishman}

\appendix
\section{Creation of the OVSA-like database of the cold solar flares}
\label{S_mw_database}

A generic OVSA data file contains low- and high- time resolution data  with the resolutions differing by a factor of two from all available antennas. There are OVSA tools developed specifically for reduction and analysis of the OVSA data.
For our data set, the input \mw\ data are, however, nonuniform as they are taken by substantially different radio instruments. To partly reduce the effect of this data nonuniformity on the statistical results we made a number of manipulations with the input data. For the events jointly observed by more than one instrument we combined the data from the different instruments into one composite dynamic spectrum to increase spectral resolution needed for meaningful time-sequential spectral fit of the data. Given the dissimilar time resolution and distinct clock ticks at various instruments, we interpolated the time array of one instrument to exactly match the clock ticks of another instrument selected to be a reference one. Whenever possible we adopted the NoRP clock ticks to be the reference ones, and resampled the RSTN and/or SRS and BBMS data to exactly match the NoRP time.

We noted that for the full NoRP time resolution in the flare mode, 0.1~s, the \mw\ light curves for many events are noisy at many frequencies. For this reason, we integrated the 0.1~s data to degrade the time resolution up to 0.5~s, which is exactly half of the NoRP resolution in the background mode. Having the two sets of observations, one with 0.5~s resolution created from the flare mode data, and the other with 1~s resolution taken from the background mode data allows us to create a data file having the structure internally identical to the generic OVSA data file. For the events for which NoRP data are unavailable (or only available in the background mode), we created a dynamic spectrum with a single time resolution, 1~s, which is also allowable by the OVSA data format.

Although a combination of more than two instruments is, in principle, possible for a given event, in practice we mainly created the composite data files containing the data from two different instruments; namely, combination of the NoRP and the RSTN data (7 events), the NoRP and the SRS data (2 events) or the NoRP and the BBMS data (1 event) and in one case the RSTN and the KMAS data, while the combination of the data from three different instruments (NoRP, RSTN, and SRS) was only available in two cases. Note, that there are often RSTN clock errors\footnote{This, in particular, implies that using RSTN data alone to measure the time delays between the X-ray and \mw\ light curves would be inconclusive.}, which can be as big as many seconds. We corrected these errors by cross-correlation between the RSTN and NoRP data, relying on the NoRP clocks as the most precise. For two events we built separate dynamic spectra from the OVSA data or the NoRP+RSTN combination and for one more---from the OVSA and the NoRP separately.  We did not add other instruments to the events observed with the OVSA, since it typically has sufficient number of the spectral channels to resolve and fit the burst spectrum without adding extra channels. However, considering a separate NoRP or NoRP+RSTN spectrum has an advantage of higher time resolution, which is important to analyze short and rapidly variable events.

\end{document}